\documentclass[trackchanges]{aastex7}
\usepackage{xcolor}
\begin{document}

\title{Extreme Geomagnetic Storm on Early Earth Induced by Corotating Interaction Regions from the Young Sun}

\author[orcid=0000-0001-9349-086X,sname='Dibyendu Sur']{Dibyendu Sur}
\altaffiliation{The Catholic University of America}
\affiliation{The Catholic University of America}
\affiliation{NASA Goddard Space Flight Center}
\email[show]{dibyendu.sur@nasa.gov}  

\author[orcid=0000-0003-4452-0588,gname=Bosque, sname='Airapetian']{Vladimir S. Airapetian} 
\affiliation{NASA Goddard Space Flight Center}
\affiliation{American University}

\email{vladimir.airapetian@nasa.gov}

\begin{abstract}

Recent multiwavelength observations of young solar analogs suggest that the young Sun in the first 600 Myr was a magnetically active star that produced an X-ray and Extreme-UV (EUV) bright corona, fast, massive stellar wind, and energetic eruptive events. These outputs affected
magnetospheric environments of early Earth and young rocky exoplanets. The interaction of the fast solar wind with the slow wind produced strong shocks from Corotating Interaction Regions (CIRs) that provided high dynamic pressure on the magnetospheres of early Venus, Earth, and Mars. Here, we apply the Space Weather Modeling Framework (SWMF), coupled with the Rice Convection Model (RCM) to simulate the response of the magnetospheric environments and associated Joule heating deposited in the upper atmosphere of early Earth as it passed through a CIR shock from the young Sun. The model suggests $\sim$ 40\% closer dayside magnetopause standoff distance, and higher negative SYMH, Cross Polar Cap Potentials (CPCP), atmospheric Joule heating, Field Aligned Currents (FAC), electron precipitations, and equatorward polar cap expansions, comparable or exceeding those of recent intense and super geomagnetic storms. The magnetic storm produces dawn-dusk asymmetries in the polar cap boundary arising from the stellar magnetic field B$_y$. The proton density enhancements during the CIR event are the dominant factor in the overall dynamic pressure for resulting electron precipitation and Joule heating deposited into the Earth's ionosphere. We discuss implications for the magnetospheric states and associated habitability conditions on early Earth and young rocky exoplanets around magnetically active solar-like stars.

\end{abstract}



\section{Introduction} 

Our planet is located in the dynamic environment of the Sun and is subject to space weather in the form of solar X-ray, Extreme UV (EUV), solar wind, Interplanetary Magnetic Field (IMF), and solar eruptive events. Specifically, energy outputs from solar flares, Coronal Mass Ejections (CMEs), and associated solar energetic particles may introduce large perturbations to the Earth's magnetosphere, ionosphere, thermosphere, and mesosphere. High-speed streams, associated Corotating Interaction Regions (CIRs), and CMEs can introduce up to two orders of magnitude of enhanced solar wind dynamic pressure and IMF that can severely affect the Earth's environment \citep{Schrijver2015}.

Recent observations of young solar analogs suggest that our Sun in its first 500 Myr of life (the young Sun) was a magnetically active star that produced hot corona, fast massive wind, and energetic flares accompanied by CMEs \citep{Airapetian2020, Namekata2024a, Namekata2024b}. This implies that the young Sun created extreme space weather conditions on the early Earth's environments at the time when life started on Earth 4.5 to 3.7 billion years ago  \citep{Rosing1999, Pearce2018}. These  extreme space weather events frequently affected the Earth's magnetospheric environments, causing disturbances or magnetic storms. With the magnetic field severely compressed by the fast and dense solar wind, the Earth encountered large atmospheric heating and polar cap expansion causing  enhanced atmospheric escape, including ion outflow \cite {Strangeway2005, Mironova2015, Airapetian2017, Airapetian2020, GarciaSage2017}.  This provided favorable conditions for solar energetic particles to penetrate the lower Earth atmosphere, ionize and dissociate atmospheric species and modify its atmospheric chemistry \citep{Airapetian2016, Kobayashi2023}. This suggests that habitability conditions could be challenged by these magnetic storms produced by massive winds and eruptive event from the active young Sun \citep{Airapetian2020}. To assess the magnetospheric and atmospheric environments of the early Earth, we need to understand how its magnetosphere responded to these storms and assess the respective impacts on the habitability.

Recent three-dimensional (3D) magnetohydrodynamic (MHD) simulations of the coronal and wind environments of ${\kappa}^1$ Ceti, the proxy star of the young Sun (age estimated to be $\sim$ 600 Myr by \cite{Airapetian2021}), output the stellar wind density $\sim$ 100 times greater and twice as fast than that of the quiet phase of the current Sun's wind at 1 AU. The interaction of a stream of high-speed solar wind originating in stellar coronal hole structures with the preceding slower wind formed along the equatorial regions of the stellar corona produces regions of compressed plasma referred to as Stream Interaction Region (SIR), along the leading edge of the stream, which, due to the rotation of the star at 9.2 days, is twisted approximately into a Parker spiral. Because the coronal holes may persist for many months, the interaction regions and high-speed streams tend to sweep past an exoplanet at regular intervals of approximately half of the stellar rotation period, forming CIRs along the spirals. Due to larger difference between the fast and slow wind components than that of the current Sun, its compression regions steepen into strong shocks propagating at $ > $ 1000 km/s at the orbital locations of Mercury (0.4 au), Venus (0.7 au), and early Earth (1 au). The wind shock densities are a factor of 300 greater than the solar wind’s density measured by Parker Solar Probe \citep{Kim2020}. This suggests that CIRs can exert dynamic pressures at $\sim$ 1300 times of the solar wind and are comparable to those formed by extreme Carrington-type CMEs. \cite{Airapetian2016} modeled the effects of such a CME event from a young Sun to the Earth's magnetosphere.

During a CIR event, increased stellar dynamic pressure pushes the planetary dayside magnetopause closer to the planet. The magnetic reconnection between stellar and planetary magnetic fields (such as the Earth) causes the opening of magnetic field lines and the expansion of the polar cap region (the region closer to the magnetic poles consisting of open magnetic field lines). The magnetic field lines reconnect again at the nightside \citep{Dungey1961}. The reconnection transfers stellar wind energy to the planetary magnetosphere. During this period, the intensity of the planetary ring current increases, which reduces the strength of the planetary magnetic field, causing a magnetic disturbance or storm. The Field Aligned Current (FAC) also increases during the magnetic storm. This increases the precipitation of charged particles into the ionosphere and subsequently the ionospheric conductances. Together with higher plasma convection, this also increases the ionospheric electric field and the Joule heating. The Joule heating causes thermospheric expansion and enhances atmospheric escape from the planet. Details of CIR events, as experienced from the Earth during the current epoch, were elaborated by \cite{Tsurutani1995, Gonzalez1999, Borovsky2006}. 

The objective of our paper is to study the response of the early Earth's magnetosphere to CIR-driven shocks from the young Sun (${\kappa}^1$ Ceti) using the Space Weather Modeling Framework (SWMF) Geospace model \citep{Toth2005, Glocer2009, Toth2012, Gombosi2021}. Our simulation also includes Rice Convection Model (RCM) \citep{Sazykin2002, DeZeeuw2004, Toffoletto2003, Wolf2007}.
We have used the CIR shock, which is a part of the solution for the stellar wind from the young Sun's analog ${\kappa}^1$ Ceti ($\sim$ 600 Myr) \citep{Airapetian2021}. This study provides insights into how far these shocks affect the magnetosphere and ionosphere of an Earth-like planet at 1 AU. This will determine their role in the evaluation of the planet's sustaining habitability and associated atmospheric escape. 

The organization of the paper is outlined below. Section 2 briefly describes the SWMF model and its subsections, as well as the the coupling mechanism of SWMF to RCM model.  Section 3 presents the MHD simulation results. Section 4 comprises the summary of the results and discussed potential limitations and the future scope of the study. We have also highlighted the implications of the model to early terrestrial (exo)planets .

\section{Model Description} 

We have performed the simulation using the SWMF, coupled with RCM
under the CCMC environment. The SWMF is a modeling framework \citep{Toth2005, Glocer2009, Toth2012, Gombosi2021} that consists of different components for different physical regions between the Sun and the Earth. The components of SWMF/Geospace are Global Magnetosphere (GM) (Block-Adaptive-Tree-Solarwind-Roe-Upwind-Scheme, BATSRUS, \citep{Powell1999, Gombosi2004}, Ionospheric Electrodynamics (IE) (Ridley Ionosphere Model (RIM), \citep{Ridley2001, Ridley2004}, Inner Magnetosphere (IM) (for example: RCM), and Radiation Belt (RB) 
. SWMF was validated to reproduce realistic space-weather parameters \citep{Rastatter2016, Liemohn2018, Sur2025}.
RIM is an electrostatic solver for the ionospheric potential given an FAC pattern and ionospheric conductance. \citep{Ridley2001, Ridley2004, Welling2019, Mukhopadhyay2020, Mukhopadhyay2022}. The ionospheric Hall and Pedersen conductances derived from this method have contributions from the solar illumination, auroral precipitation, and other sources such as polar rain and starlight. Solar illumination-driven conductance is calculated using inputs from the F10.7 index (a proxy for Extreme UV emission) and solar zenith angle \citep{Moen1993}. Conductance due to auroral precipitation is computed using an empirical relation with FAC as input. FAC maps were computed from several ground magnetometer measurements during January 1997 using the Assimilative Mapping of Ionospheric Electrodynamics (AMIE) method \citep{Richmond1988}. The relation depends on latitudes and longitudes for upward and downward FAC. 

RCM was developed at Rice University to characterize magnetospheric and ionospheric electrodynamics self-consistently \citep{Wolf1970, Harel1981, Sazykin2000} inside a closed magnetic field line configuration. The region generally covers the dayside magnetopause to the half of the nightside plasma sheet ($\sim$30R$_E$) \citep{Toffoletto2003, DeZeeuw2004}. RCM provides magnetospheric ion and electron distributions for energy up to a few hundred keV \citep{Wolf1982, Yang2008, Hudson2014, Sun2025a, Sun2025b}. RCM was extensively used for different space weather applications \citep{Spiro1981, Wolf1982, Fejer1990, Zhang2009, Priyadarshi2021} to name a few.


In SWMF, BATSRUS and RCM are two-way coupled to provide the realistic interaction between the ionosphere and the magnetosphere \citep{DeZeeuw2004}. BATSRUS solves three-dimensional MHD equations under finite volume to produce magnetospheric configurations. It provides the dynamic magnetospheric configurations to the two-dimensional inner magnetospheric models, such as RCM,
by closed magnetic field-line tracing, and after every specific time interval. It also provides the boundary conditions for electric potentials and plasma distributions to RCM. IE solves the ionospheric potential and provides it to RCM after interpolating into RCM grids. RCM provides realistic inner magnetospheric ring current plasma pressure back to BATSRUS. Including RCM in the SWMF ensures realistic Region-2 FAC information is obtained. The details of BATSRUS-RCM coupling were elaborated in \cite{DeZeeuw2004}. 
Overall, SWMF with RCM
accepts solar wind and Interplanetary Magnetic Field (IMF) inputs to provide comprehensive information regarding the coupled magnetosphere and ionosphere. 

The present simulation uses SWMF version 20180525 with the Space Weather Prediction Center (SWPC) real-time grid consisting of 1,007,616 cells. The simulation box size is $-224\,R_E < X < 32\,R_E$ and $|Z|, |Y| < 128\,R_E$ in GSM coordinates. The SWMF adaptive grid comprises blocks of constant spacing in all directions. It has the finest resolution of $0.125\,R_E$ near the inner boundary. At the dayside magnetopause ($X = 10\,R_E$, $Y = 0$, $Z = 0$), the resolution is $0.5\,R_E$ in all directions. The grid becomes coarser farther away from the inner magnetosphere. In the magnetotail region ($-32\,R_E < X < -16\,R_E$, $Y = 0$, $Z = 0$), the resolution is $1\,R_E$ in all directions. It should be noted that the spatial resolution had to be limited, as the model run terminated abruptly several times when using higher resolutions. CCMC-based SWMF ionospheric outputs are provided in solar magnetic coordinates.

\section{Results: Extreme Magnetic Storm on Early Earth} \label{sec:style}

\begin{figure*}[ht!]
\centering
\includegraphics[width=\textwidth,height=0.9\textheight,keepaspectratio]{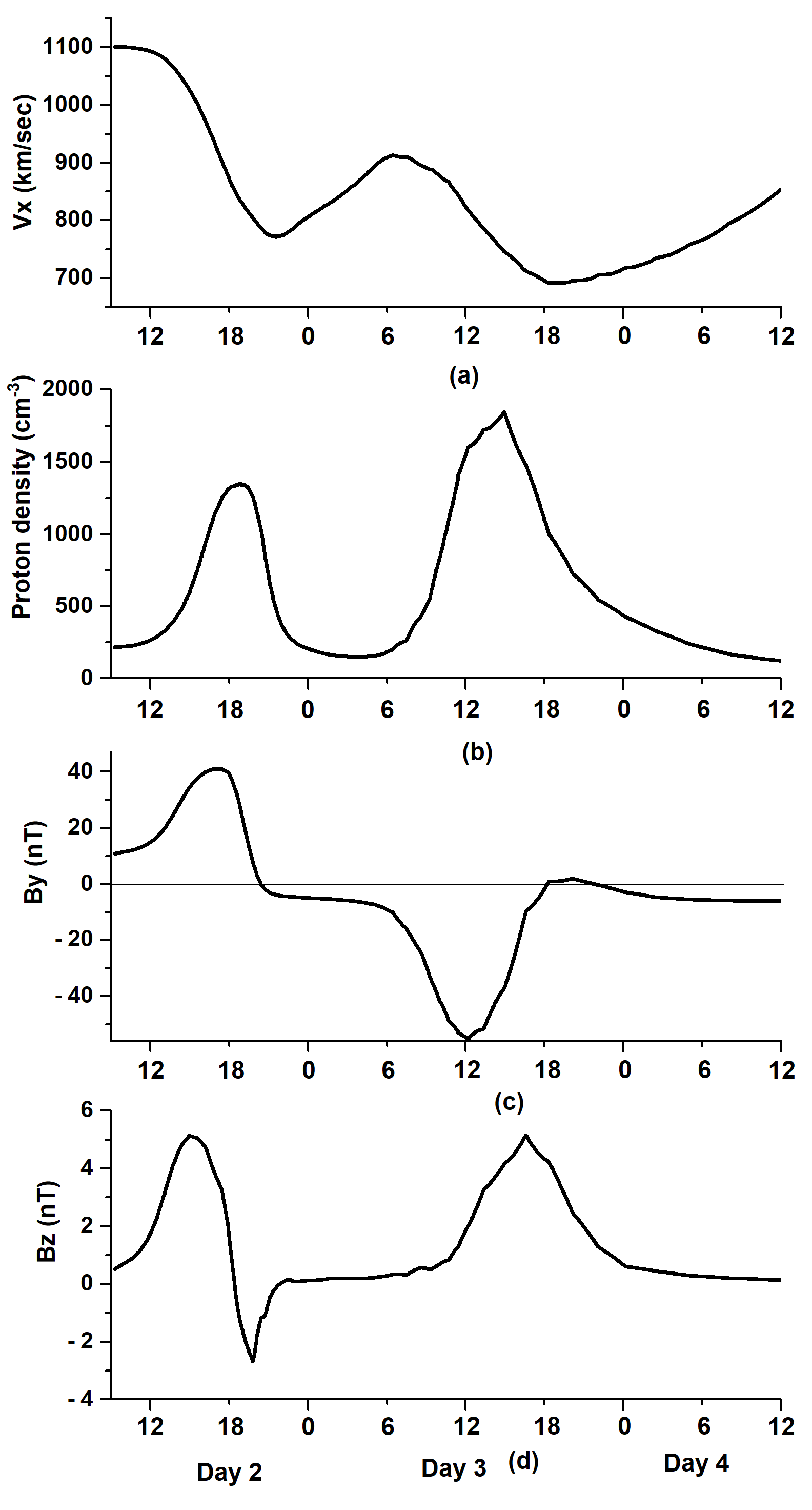}
\caption{Temporal variations of (a) stellar wind speed (V$_x$), (b) proton density, (c) B$_y$, and (d) B$_z$ component of stellar magnetic field 
\label{fig:general}}
\end{figure*}

Here, we have applied SWMF to model a specific time period from the stellar wind structure from the young Sun (see Figure 8 of \cite{Airapetian2021}). This time period covers two peaks of proton density as the Earth passes through a shocked region introduced by a CIR event from the young Sun over Days 2-4, as shown in the same figure. January 22, 2011, is labeled as Day 2, January 23, 2011, as Day 3, and January 24, 2011, as Day 4. The time period of the wind simulation used in this study spans between Day 2 (9:17 UT, January 22, 2011) and Day 4 (12:00 UT, January 24, 2011). Figure 1 shows the stellar wind speed, V$_x$ (km/sec), proton density (number/cm$^{3}$), and stellar magnetic field B$_y$, and B$_z$ (in nT) for the specified period. The CIR compressed region consists of three peaks of stellar wind speed (maximum: 1100.90 km/sec), two peaks of proton density (maximum: 1844.31/cm$^{3}$), and two peaks of dynamic pressure (maximum: 2145.17 nPa) (Figure 2). It should be noted that the maximum stellar wind speed from the young Sun during this period \citep{Airapetian2021} falls within the ranges of those observed in CME driven superstorms on the current Earth \citep{Skoug2004, Tsurutani2006}. However, the maximum proton density obtained in the MHD model is $\sim$ 18 times higher than that observed from the current transient events \citep{Skoug2004}. This shows that proton density is the primary contributor to the dynamic pressure. We have calculated the Pearson correlation coefficient (r) between (i) stellar wind dynamic pressure and proton density, and (ii) stellar wind dynamic pressure and stellar wind speed. Pearson correlation coefficient (r) has been used to observe possible correlations between two variables \citep{Liemohn2021}. This is defined in Equation 1.

\begin{equation}
r = \frac{\sum_{i=1}^{n} (x_i - \bar{x})(y_i - \bar{y})}{\sqrt{\sum_{i=1}^{n} (x_i - \bar{x})^2} \sqrt{\sum_{i=1}^{n} (y_i - \bar{y})^2}}
\end{equation}                                     

Here, x and y are the variables with n number of values in their datasets. $\bar{x}$ and $\bar{y}$ are the mean or average values for the x and y datasets, respectively.  The correlation coefficient $r \geq 0.7$ is considered to be a good correlation \citep{Liemohn2021}. The correlation between stellar wind speed and dynamic pressure yields the value of 0.04, indicating poor correlation. However, the correlation between proton density and dynamic pressure is 0.94, which indicates a strong correlation. This indicates that the variation of proton density are wider than the stellar wind speed, making it the dominant contributor to the dynamic pressure variation in the present simulation.
The amplitude of variation of stellar magnetic field B$_y$ during the simulation period (-55.31 to 41.08 nT) is wider than the corresponding B$_z$ component. We have also studied the effects of stellar magnetic field B$_y$ asymmetries in this paper.

\begin{figure*}[ht!]
\centering
\includegraphics[width=\textwidth,height=0.9\textheight,keepaspectratio]{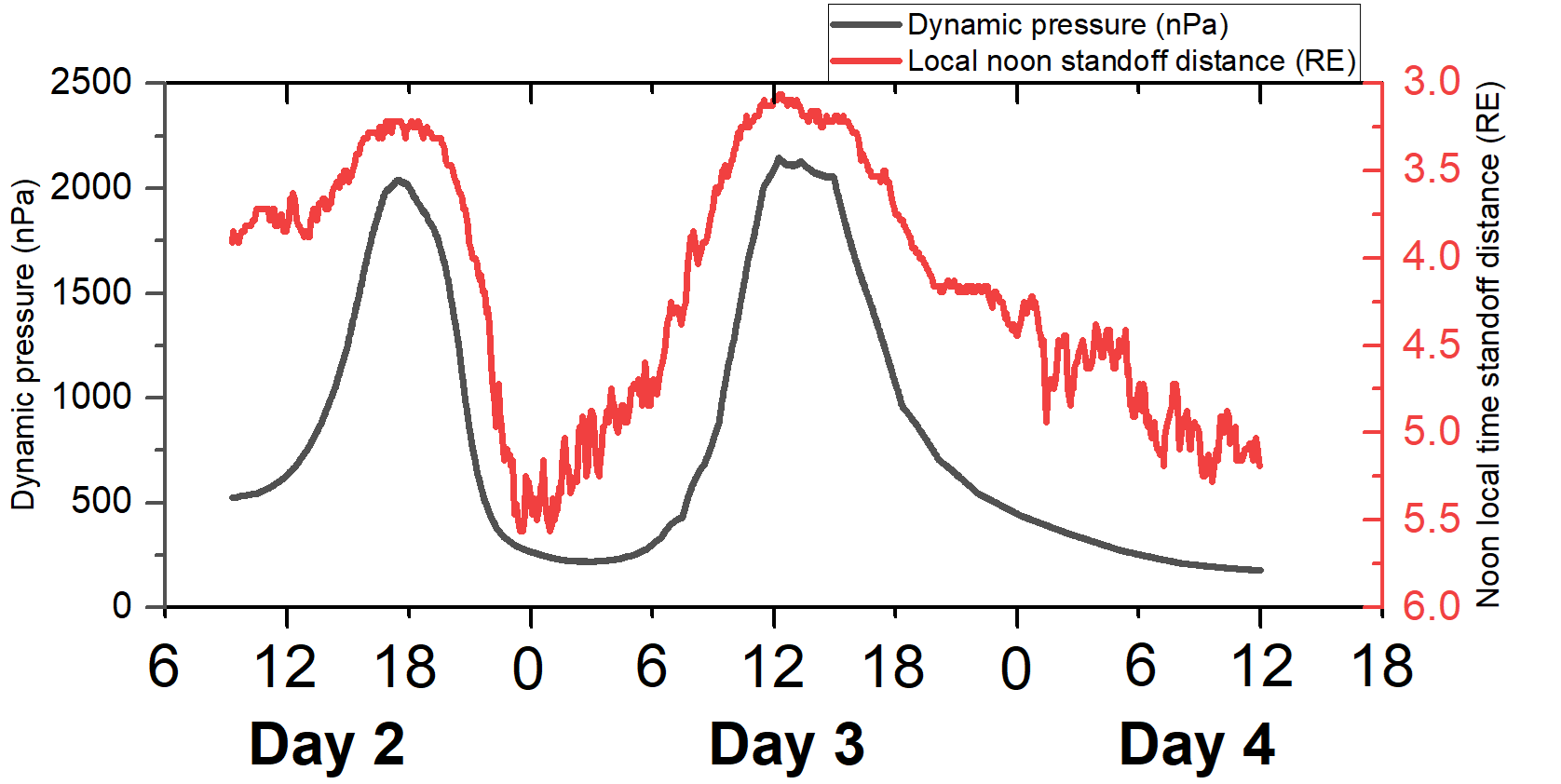}
\caption{Correlation of dynamic pressure and local noon magnetopause standoff distance
\label{fig:general}}
\end{figure*}

The minimum local noon magnetopause standoff distance from our simulation is 3.06 R$_E$ (Figure 2). This is closer to the Earth than the magnetopause standoff distance observed during May 10-11, 2024 geomagnetic storm (5 R$_E$) \citep{DaSilva2025, Fu2025}. This indicates that the CIR storm simulated in this paper compressed the Earth’s dayside magnetopause more significantly and made it $\sim$ 40\% closer to the planet. The standoff distance is also correlated with dynamic pressure, as expected (Figure 2). However, the dayside magnetopause standoff distance during the CIR event from the early Earth was farther than the same during a CME event (magnetopause standoff distance: 1.5 R$_E$) \citep{Airapetian2016}, indicating lesser intensity during the present CIR event than the CME event.

\begin{figure*}[htbp]
\centering
\includegraphics[width=\textwidth,height=0.9\textheight,keepaspectratio]{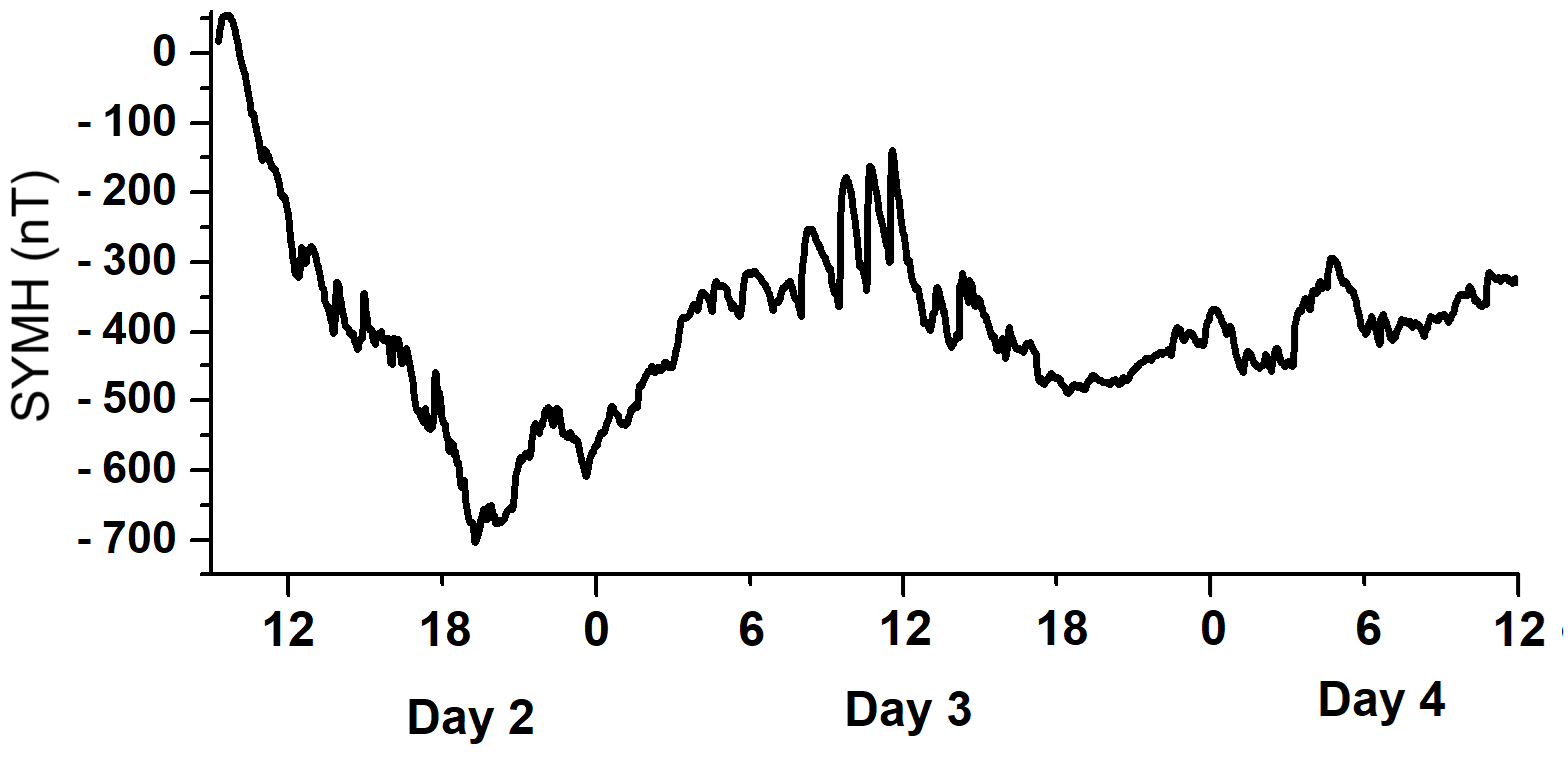}
\caption{SYMH as obtained from SWMF simulation
\label{fig:general}}
\end{figure*}

SWMF simulated Symmetric H-component (SYMH) index, Cross Polar Cap Potential (CPCP) (from both hemispheres), hemispherically integrated Joule heating rates (also from both hemispheres), and northern hemispheric polar cap boundary are shown in Figures 3, 5, 7, and 8, respectively. The storm-time strength of the ring current is measured by the Disturbed Storm Time (Dst) or Symmetric H-component (SYMH) index at hourly and one-minute resolutions, respectively. On Earth, these are calculated from horizontal components of Earth's magnetic field at the low-latitude ground-based magnetometer stations. The maximum negative SYMH obtained from SWMF simulation during this period is -703.90 nT at 19:18 UT, Day 2, indicating a super magnetic storm (Figure 3). This is higher than almost all recorded severe geomagnetic storms on the Earth since the International Geophysical Year (IGY) \citep{Stanislawska2018}, including the geomagnetic storm during May 10-12, 2024 (maximum negative SYMH = -518 nT and maximum negative Dst = -406 nT). However, SYMH from the CIR event presented in this paper is comparable with the March 13-14, 1989 geomagnetic storm (maximum negative SYMH: -720 nT). This is also weaker than the Carrington geomagnetic storm event on September 1-2, 1859 (maximum negative Dst $\sim$ -1760 nT) \citep{Tsurutani2003}. However, it must be noted that during extreme magnetospheric compression (local noon magnetopause standoff distance $\sim$ 3 R$_E$), Dst 
could be attenuated by the magnetopause and cross-tail currents. We have a reasonable spatial distance between the MHD inner boundary and the dayside compressed magnetopause in our simulation. This has been illustrated by a representative case of magnetospheric current density distribution at the Z=0 plane during 14:57 UT, Day 3  (Figure 4a). We have chosen this period since the dynamic pressure and proton density at this time instant are higher (dynamic pressure: 2053.55 nPa, proton density: 1844.31/cm$^{3}$, magnetopause standoff distance: 3.22 R$_E$). The figure shows the separation between the dayside magnetopause and the MHD inner boundary. However, this does not completely mitigate the possibility of contributions from other current sources towards the 
Dst, especially during high dynamic pressure periods. It is already established that the Dst is also affected by dayside magnetopause current and cross-tail current, apart from the ring current \citep{Turner2000}. All of these currents are also increased during the high dynamic pressure periods \citep{Rufenach1992}. \cite{Blake2021} indicated that the contributions from magnetopause current and Region 1 FAC towards Dst are comparable to those from the ring current during extreme conditions. \cite{Thomas2024} also emphasized the roles of FAC, and magnetospheric and ionospheric currents towards storm-time magnetic field depressions. The compression of the magnetopause may lead to higher plasma injection into the ring current, increasing its strength. The relation between the ring current energy and the corresponding magnetic depression is linear or nearly  linear \citep{Dessler1959, Sckopke1966, Russell1997, Turner2001}. Using this assumption, we have calculated magnetic depression due to the RCM ring current. Figure 4b presents the temporal variations of RCM ring current driven SYMH alongside the total SYMH. 
The magnitude of total SYMH is higher than the ring current driven SYMH, especially near the first peak of the dynamic pressure and negative stellar magnetic field B$_z$. For example, we found that the magnetic depression due to RCM ring current at 19:17 UT, Day 2 is $\sim$ -513.71 nT. Corresponding total SYMH from the simulation is -703.17 nT. This suggests that a significant amount of additional magnetic depression (189.46 nT) may come from other current sources. The other possible current sources are discussed in \citep{Liemohn2003, Blake2021, Thomas2024} and are also mentioned above. In addition, Figure 4a also shows the dawn-dusk asymmetry in tail current, possibly due to the stellar magnetic field B$_y$ asymmetry, consistently with \cite{Cheng2014, Sibeck2014, Zhang2023}.

\begin{figure*}[htbp]
\centering
\includegraphics[width=\textwidth,height=0.9\textheight,keepaspectratio]{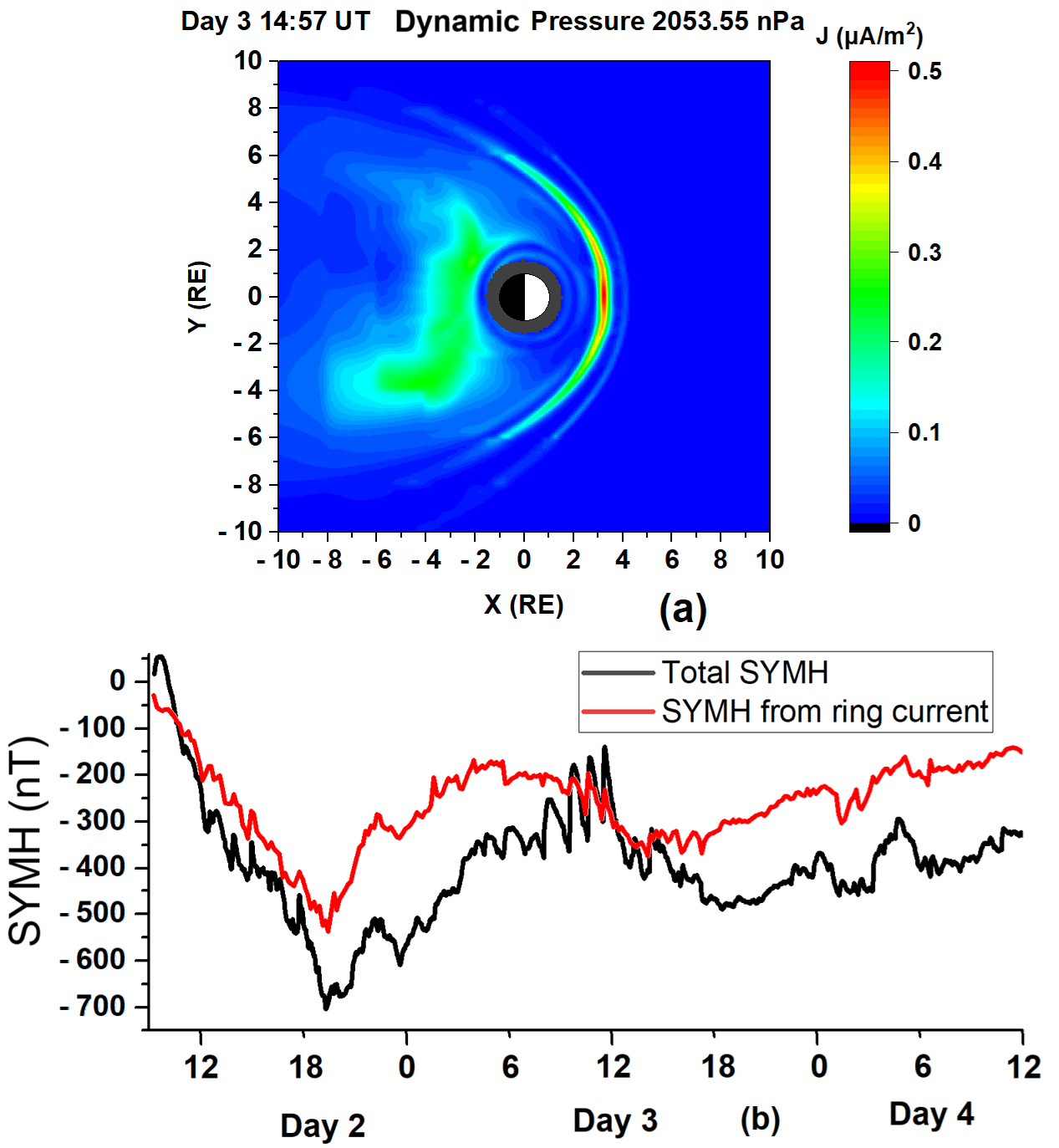}
\caption{ (a) SWMF current density in the planetary magnetosphere at the XY plane (b) Temporal variations of total SYMH and RCM ring current driven SYMH 
\label{fig:general}}
\end{figure*}

In the present case, the variation of stellar magnetic field B$_z$ is narrower than B$_y$ variations (B$_z$ variation ranges between -2.69 nT and 5.15 nT). 
It should be noted that, short duration or low-magnitude southward IMF B$_z$ may also explain some geomagnetic storm events. 
An example of short duration southward IMF B$_z$ was the geomagnetic storm event during January 21-22, 2005 \citep{Du2008, DU201255}. 
On January 21, 2005, IMF B$_z$ was southward and fluctuating for two short intervals ($\sim$ 30 minutes each and reaching $\sim$ -10 nT and $\sim$ -26 nT, respectively). These produced small negative SYMH excursions.  
After that, IMF B$_z$ became northward during the main phase of the storm, with a maximum negative SYMH of $\sim$ -100 nT, which occurred several hours after the storm began. Higher values in the solar wind speed ($\sim$ 900 km/sec) and dynamic pressure (peak $\sim$ 90 nPa) were also seen during that period. In our case, the stellar magnetic field B$_z$ was southward for three and a half hours and was non-fluctuating. The dynamic pressure in our case is around 20 times higher. \cite{Du2008} suggested that this event can be described by the transfer of solar wind energy to the magnetotail, which was transferred to the magnetosphere and strengthened the ring current after a time lag. They included viscous interactions between the solar wind and the Earth's magnetosphere, and the influence of cross-tail current on SYMH as other possible explanations. They also suggested that other explanations are also possible. 
For example, \cite{Dmitriev2014} showed that the magnetopause standoff distance was 3 R$_E$ during that event. The paper suggested that the compressed magnetopause pushed the outer radiation belt and ring currents inward towards the Earth (L $ < $ 4) and accelerated ring current particles. 
This possibly increased the ring current strength and reduced the Dst. The case we presented in this paper is also similar, as the magnetopause standoff distance is similar, and the ring current decay is also slower. \cite{Newell2007} also indicated that as the magnetopause comes closer to the ring current during higher dynamic pressure, it may also affect the overall strength of the ring current. 

\cite{Troshichev2011a, Troshichev2011b} emphasized the contributions from IMF B$_y$ towards the geoeffective interplanetary electric field (E\textsubscript{geoeff}) \citep{Kan1979}. The geoeffective interplanetary electric field is defined in Equation 2.

\begin{equation}
E_{\text{geoeff}} = V \left( B_Y^2 + B_Z^2 \right)^{1/2} \sin^2 \left( \frac{\phi}{2} \right)
\end{equation} 

Here, V is the solar wind speed, B$_y$ and B$_z$ are the IMF components, and $\phi$ is the angle between the Earth's geomagnetic dipole and the transverse IMF. $E_{\text{geoeff}}$ determines the strength of solar wind and the Earth's magnetospheric coupling. The relation indicates the significant contributions from IMF B$_y$, especially when the magnitude of IMF B$_z$ is small and northward \citep{Troshichev2011b}. As in our case, the magnitudes of the stellar magnetic field B$_y$ were higher than the stellar magnetic field B$_z$. This can explain the stronger magnetic reconnection and the corresponding greater negative SYMH, as seen from Figure 3. With a higher transverse IMF component, the necessity of southward IMF B$_z$ for magnetic reconnection is reduced \citep{Gonzalez1974}. Possible reconnection during northward IMF alongside higher IMF B$_y$ component was also supported by \cite{Kuznetsova2011}. \cite{Newell2007} summarized several solar wind-magnetosphere coupling functions, consisting of the IMF clock angle and total IMF magnitude, both of which vary with IMF B$_y$. The clock angle is crucial for magnetic reconnection \citep{Fedder1991} and determines the location of the reconnection line \citep{White1998, Petrinec2016, Fuselier2024} 

Therefore, the greater negative SYMH observed in our case may be caused by the strong dayside stellar wind-magnetosphere coupling due to geoeffective interplanetary electric field, mainly driven by the stellar magnetic field B$_y$ and high stellar wind speed. This could also be supported by the approximately three and a half hours southward stellar magnetic field B$_z$ (even with a lower magnitude) to initiate the process. This could be observed in Figure 3 showing two peaks of the negative SYMH  (first peak at 19:18 UT, Day 2, and second peak at 18:27 UT, Day 3). These peaks are closely correlated with the dynamic pressure, indicating its influence. However, the first peak at 19:18 UT, Day 2, has a higher negative magnitude. This can be explained by the contribution from the southward stellar magnetic field B$_z$ during this period. The role of dynamic pressure in reconnection and strength in the ring current can be explained mainly by the close vicinity of the magnetopause to the inner magnetosphere region, which can increase the ring current.
The contributions from other sources to Dst or SYMH, especially during compressed magnetosphere periods due to extreme dynamic pressure, may also explain the higher negative SYMH seen in our simulation. However, this makes Dst unreliable for determining the geomagnetic storm intensities. Therefore, although we have shown the comparison between Dst/SYMH values from different geomagnetic storms and our simulation above, this is not a direct indication of the contribution of the ring current strength to the geomagnetic storm intensity \citep{Blake2021}. 

\begin{figure*}[htbp]
\centering
\includegraphics[width=\textwidth,height=0.9\textheight,keepaspectratio]{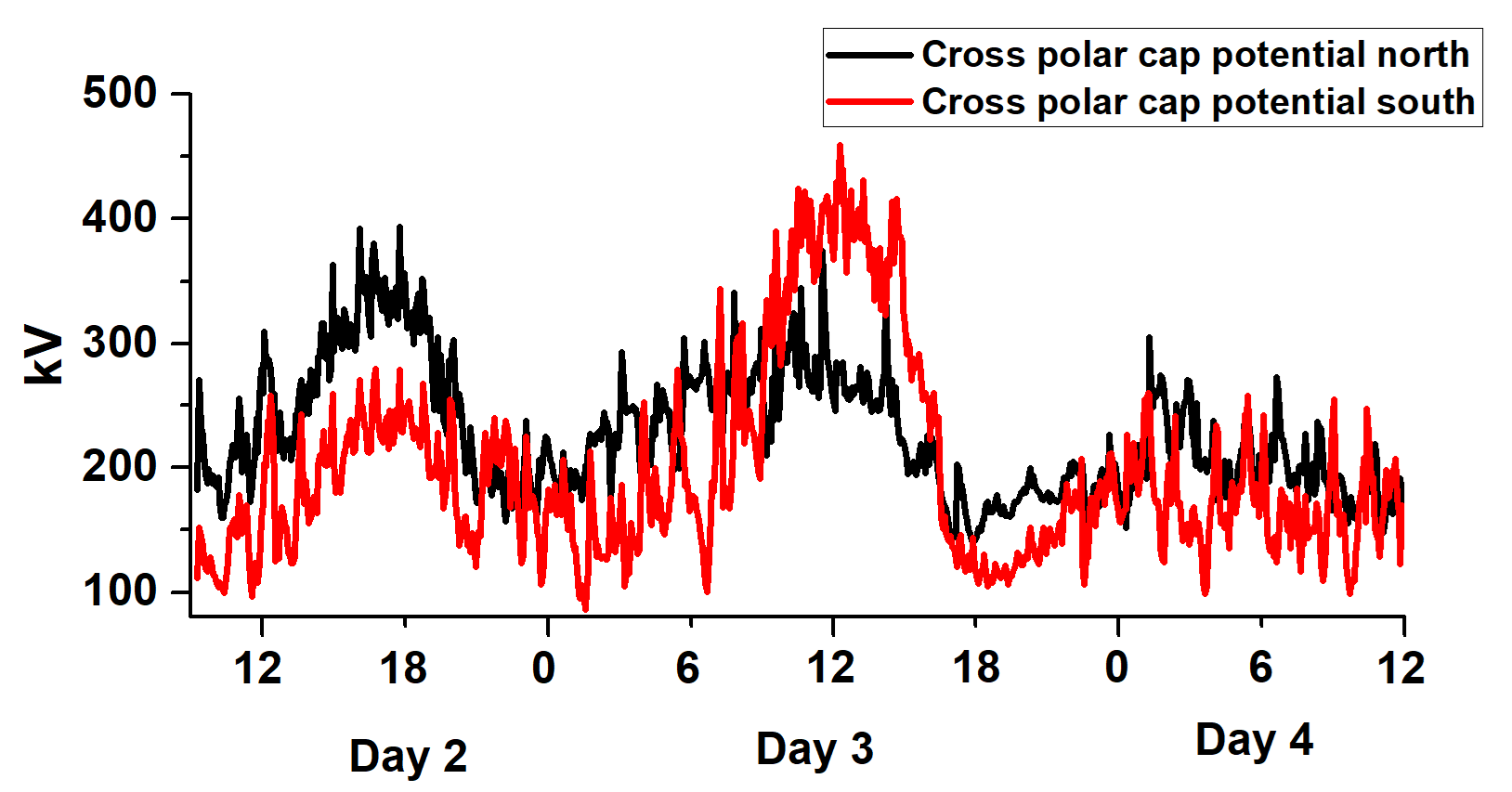}
\caption{CPCP as obtained from SWMF simulation
\label{fig:general}}
\end{figure*}

\begin{figure*}[htbp]
\centering
\includegraphics[width=\textwidth,height=0.9\textheight,keepaspectratio]{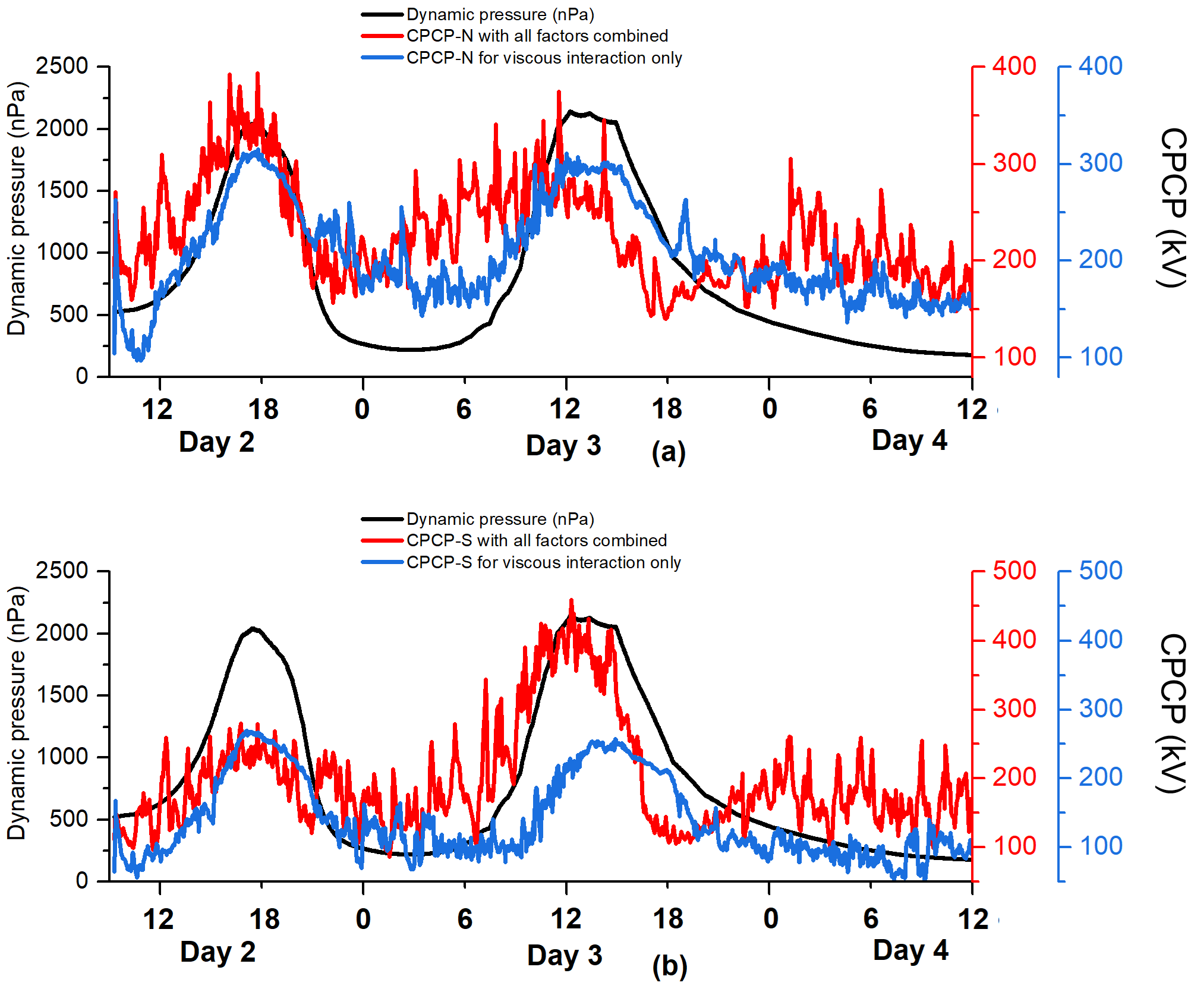}
\caption{Temporal variations of CPCP from main simulation and simulation2 (from viscous interaction only) alongside the dynamic pressure
\label{fig:general}}
\end{figure*}

During a CIR event, the transfer of stellar wind energy to the planetary magnetosphere increases the plasma convection and the potential differences across the polar cap. This is referred to as CPCP, which is measured as the difference between maximum and minimum potentials in the high latitudes for one hemisphere. CPCP from SWMF simulation shows two peaks on Days 2-3, alongside one minor peak on Day 4 (Figure 5). During the first peak, CPCP from the northern hemisphere is higher than that from the southern hemisphere. The opposite is seen during the second peak. The peak CPCP values are 393.34 kV and 458.83 kV from the northern and southern hemispheres, respectively. This is comparable to the same during the Halloween storm (October 29, 2003, maximum negative Dst = -350 nT), as obtained from AMIE and a three-dimensional global multi-fluid model \citep{Harnett2008}. The smoothed CPCP values from AMIE and the multi-fluid models were 300 kV and 400 kV, respectively, with the latter showing short-duration peak values of $\sim$ 600 kV. CPCP $\sim$ 300 kV were also detected from Defense Meteorological Satellite Program (DMSP) satellite observations during October 29-31, 2003 (maximum negative DST = -383 nT) \citep{Huang2007}.

\begin{figure*}[htbp]
\centering
\includegraphics[width=\textwidth,height=0.9\textheight,keepaspectratio]{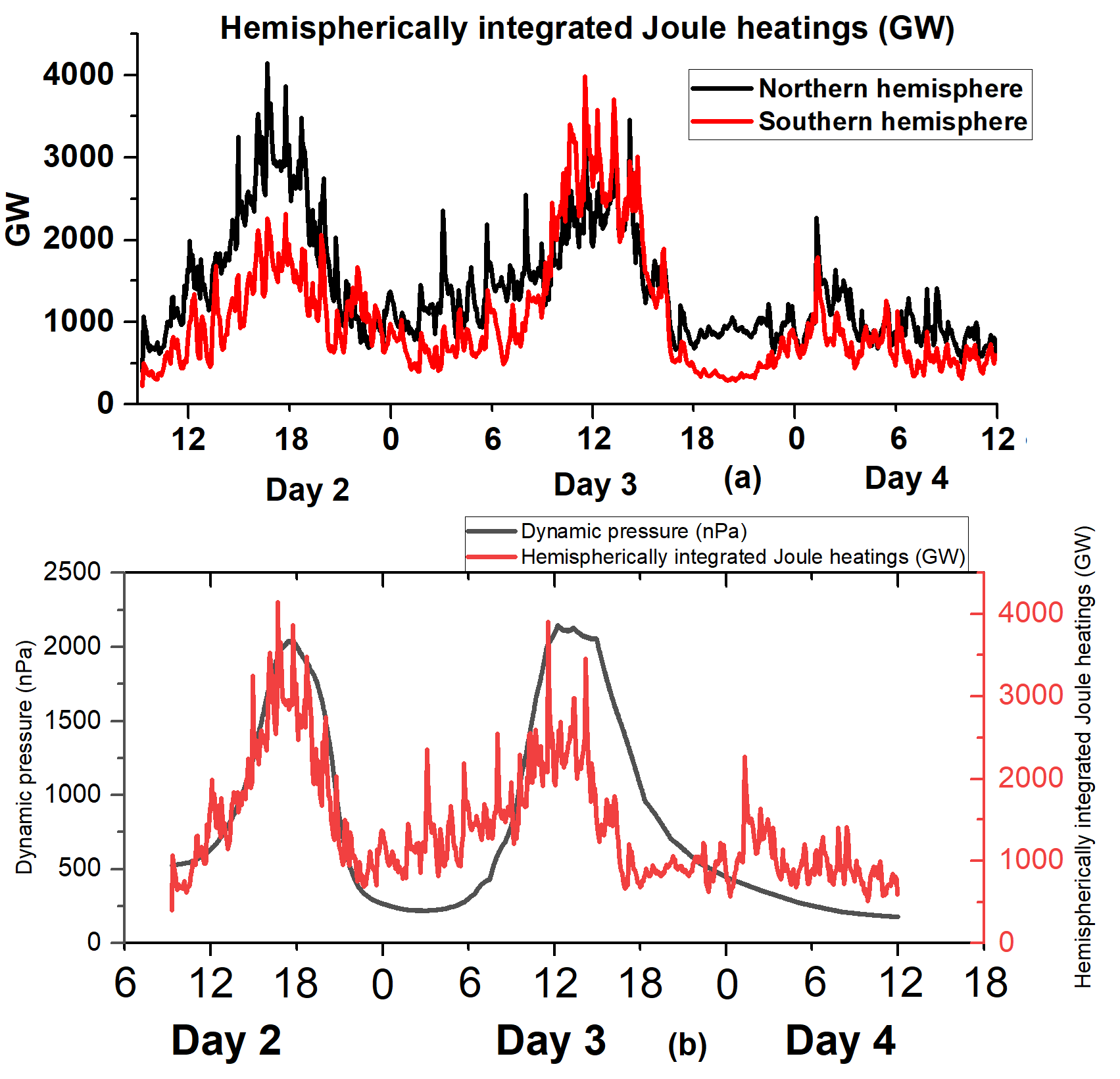}
\caption{(a) Hemispherically integrated Joule heating rates as obtained from SWMF simulation, and (b) Correlation of dynamic pressure and northern hemispherically integrated Joule heating rates
\label{fig:general}}
\end{figure*}

However, the higher CPCP during the present storm period can be explained by the high stellar wind speed and the presence of three and a half hours of southward stellar magnetic field B$_z$, which supported the transfer of stellar energy to the magnetosphere. Moreover, the solar wind-magnetosphere coupling and subsequent transfer of the solar wind energy can also be explained by the geoeffective interplanetary electric field ($E_{\text{geoeff}}$, Equation 2). We have found good correlations (Pearson's r coefficient) between $E_{\text{geoeff}}$ and CPCP (North and South), 0.73 and 0.77, respectively. This includes contributions from the B$_y$, B$_z$ components of the stellar magnetic field, and the high stellar wind speed. While coupling through the geoeffective interplanetary electric field is one of the driving factors for higher CPCP, viscous interactions between the solar wind and the Earth's magnetosphere also possibly play an important role, especially when IMF B$_z$ is very low or northward. \cite{AxfordHines1961, Axford1964} indicates that high solar wind speed, dynamic pressure, and proton density, all contribute to the viscous interaction and solar wind energy transfer to the magnetosphere. We have  
inspected the contribution of dynamic pressure to CPCP using Pearson's r coefficients of 0.55 and 0.67 for CPCP North and CPCP South, respectively. This indicates that the dynamic pressure also contributes to high CPCP, especially through the viscous interactions between solar wind and the Earth's magnetosphere. \citep{Boudouridis2007} showed that higher dynamic pressure increases dayside magnetospheric reconnection. They also showed that the dynamic pressure increases ionospheric plasma convection, both during northward and southward IMF B$_z$ conditions. Moreover, the proton density, which is nearly correlated with the dynamic pressure and contributes to the viscous interaction, has correlation coefficients of 0.33 and 0.61 with CPCP from both hemispheres, respectively. The correlation between proton density and CPCP is consistent with \cite{Ober2006}.

In order to investigate the role of viscous interaction between the stellar wind and Earth's magnetosphere during the absence of the Dungey reconnections, we have performed another SWMF+RCM simulation (simulation2) with a similar setup and the solar wind inputs as for the main simulation. Here,  we set the IMF B$_y$ = 0 and IMF B$_z$ positive. The stellar wind speed and proton density values are kept the same as in the original simulation. The temporal variations of CPCP values from the main simulation (all factors such as Dungey magnetic reconnection and viscous interaction combined, in red) and simulation2 (viscous interaction only, in blue) are shown in Figure 6. The results from the northern and southern hemispheres are shown in Figures 6a-b, respectively. The dynamic pressure is also shown alongside the CPCP variations. The maximum CPCP values derived from simulation2 (viscous interaction only) are 315.10 and 268.51 kV from the northern and southern hemispheres, respectively. These are lower than the maximum values from the main simulation (393.34 kV and 458.83 kV from the northern and southern hemispheres, respectively). However, simulation2 values are comparable with respect to the main simulation results. It should be noted that this is a separate and independent simulation. Each simulation consists of different and nonlinear MHD responses and nonlinear coupling with stellar wind drivers. Therefore, the contributions from Dungey reconnection and viscous interaction to CPCP are not strictly additive from two separate MHD simulations. However, it can be seen that the CPCP values from simulation2 are more in correlation with the dynamic pressure than the main simulation, indicating the contributions from the viscous interaction. We have found that the Pearson r coefficients for dynamic pressure are 0.89 and 0.90, respectively, from the two hemispheres for simulation2 (the same from the main simulation are 0.55 and 0.67, respectively, as mentioned above). The correlations with the proton density are also higher in simulation2 (0.87 and 0.90, respectively) than in the main simulation (0.33 and 0.61, respectively). 
This indicates moderate but comparable contributions from viscous interactions to the CPCP due to high dynamic pressure, proton density, and stellar wind speed in the present simulation, compared to the reconnection-driven interaction. 

\begin{figure*}[htbp]
\centering
\includegraphics[width=\textwidth,height=0.9\textheight,keepaspectratio]{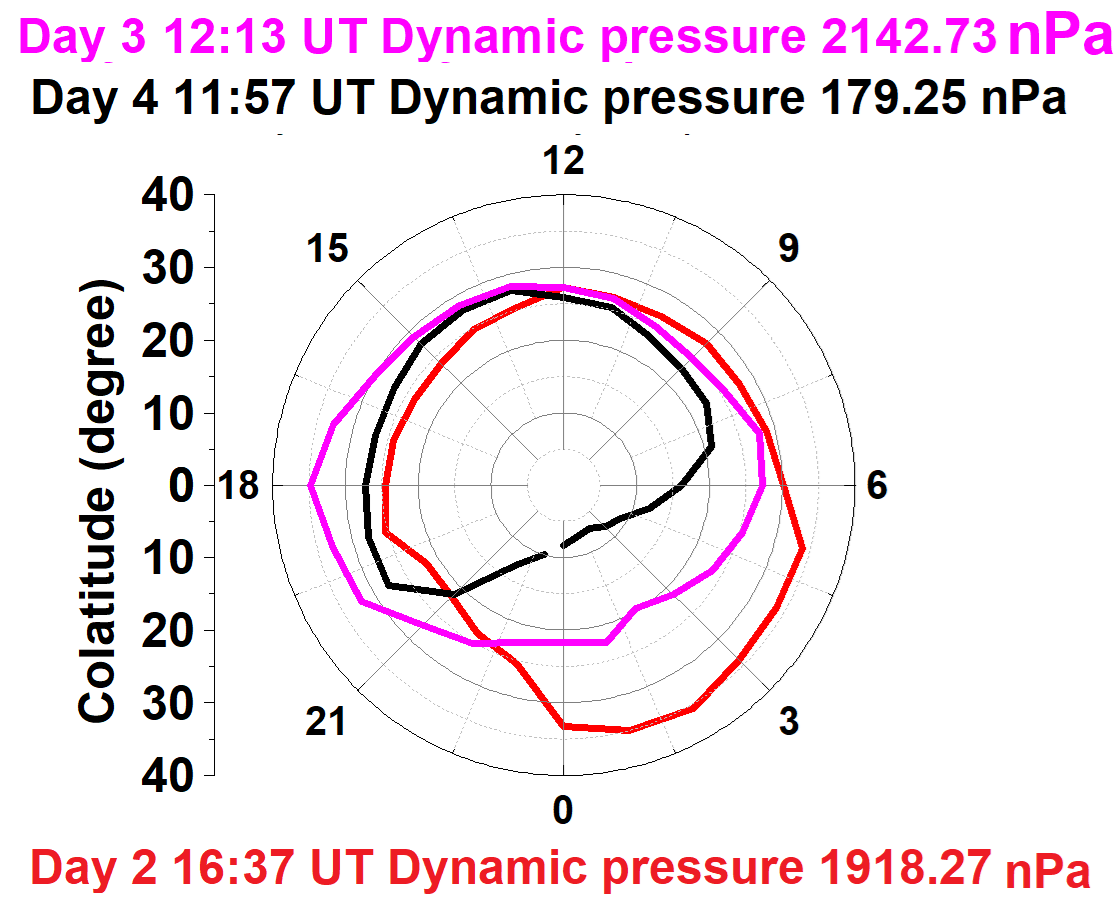}
\caption{Northern hemispheric polar cap boundary during lower and higher dynamic pressure and positive and negative stellar magnetic field B$_y$ periods, as obtained from SWMF simulation
\label{fig:general}}
\end{figure*}

Figure 7a shows high hemispherically integrated Joule heating rates (maximum: 4145.25 GW and 3981.6 GW from the northern and southern hemispheres, respectively). These also show two prominent peaks and one minor peak. During the first peak, the values from the northern hemisphere are higher than those from the southern hemisphere. The opposite is true during the second peak, consistent with the respective CPCP values. We have compared the historical total radiative cooling rates from NO and CO$_2$ combined with our model-simulated storm-time Joule heating rates, assuming they are comparable, though acknowledging that other processes are also involved. We have found that the daily Joule heating rate power values from one hemisphere from our simulation are comparable to the sum of total daily global power emitted due to NO and CO$_2$ radiative cooling, as obtained from historical observations (January 22, 2002, to May 28, 2024) \citep{Mlynczak2024}, including the disturbed periods. The average daily power values of Joule energy deposition from the northern hemisphere on Days 2-4 are 1.77 TW, 1.41 TW, and 0.97 TW, respectively, from our study. The total from these three days is 4.15 TW. The same from the southern hemisphere are 1.10 TW, 1.18 TW, and 0.69 TW, respectively, with a total of 2.98 TW. The total powers from each hemisphere are higher than the sum of global observed NO and CO$_2$ cooling power above the background during May 10-13, 2024 (2.64 TW), as seen from the NASA Thermosphere‐Ionosphere‐Mesosphere Energetics and Dynamics satellite (TIMED) Sounding of the Atmosphere using Broadband Emission Radiometry (SABER) data \citep{Mlynczak2024}. The corresponding powers of Joule energy dissipation from individual hemispheres and individual disturbed days (Day 2, 3, or 4 of our simulation) are also similar to or higher than the individual day total global radiated cooling powers as seen on May 10-13, 2024. The global energy radiated during the four days (May 10-13, 2024) due to NO and CO$_2$ cooling was 2.28×10\textsuperscript{17} Joules. The values of energy dissipated due to Joule heating from our simulation are 2.57×10\textsuperscript{17} Joules and 1.90×10\textsuperscript{17} Joules for northern and southern hemispheres, respectively. Moreover, a superstorm comparable to the May 10-13, 2024 storm typically occurs on average once in every 11 years or $\sim$ 0.1\% of the time interval. In contrast, early Earth would pass through 3-4 CIR events per stellar rotation period \citep{Airapetian2021} or spend $\sim$ 11-14\% of the total time interval. Such frequent interactions with very high stellar energy input could substantially impact the chemistry of the planetary atmosphere, a critical factor for its habitability. 
Figure 7b shows good correlations between the dynamic pressure and the simulated northern hemispherically integrated Joule heating rates, especially during the first peak. This is consistent with \cite{Palmroth2004}.

\begin{figure*}[htbp]
\centering
\includegraphics[width=\textwidth,height=0.9\textheight,keepaspectratio]{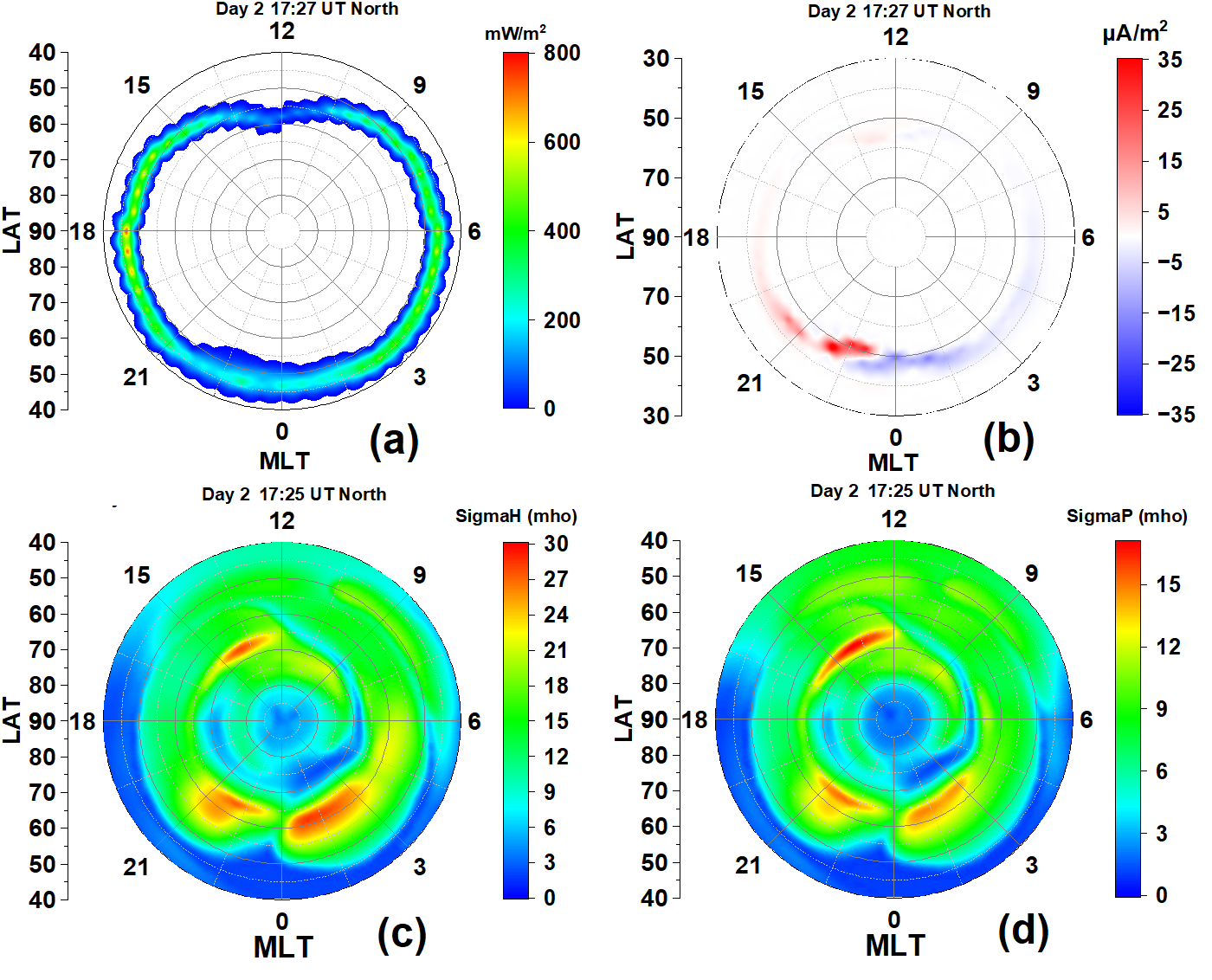}
\caption{(a) Integrated energy flux from SWMF (RCM) electron precipitation in the northern hemisphere during high dynamic pressure period. Corresponding FAC (b), and ionospheric height-integrated Hall and Pedersen conductances (c-d) are also shown.
\label{fig:general}}
\end{figure*}

Figure 8 shows northern hemispheric polar cap boundaries (colatitudes) obtained during relatively higher (16:37 UT, Day 2) and lower (11:57 UT, Day 4) dynamic pressure periods. We have found that the polar cap boundary extends to relatively lower latitudes for a higher dynamic pressure period, especially during nighttime (maximum colatitude 35.51° at 2 MLT). The distributions of the polar cap region also show dawn-dusk asymmetries, possibly due to different polarities of the stellar magnetic field B$_y$. Stellar magnetic field B$_y$ during 16:37 UT, Day 2 was 40.72 nT (positive), pushing the polar cap dawn side. In contrast, during 11:57 UT, Day 4, B$_y$ was -5.83 nT (negative), which pushed the polar cap boundary duskward. We have shown another representative case of higher dynamic pressure and negative stellar magnetic field B$_y$ at 12:13 UT, Day 3 (B$_y$: -55.29 nT). The expansion of the polar cap for this case is higher than that at 11:57 UT, Day 4. The expansion is mostly at the dusk side. These results are consistent with \cite{Akasofu1991, Elphinstone1993, Trondsen1999, Lukianova2011}. Peak northern hemispheric polar cap extension is observed during 1:37 UT, Day 3. During this period, the polar cap expands to 41.28° colatitude towards dusk for a very brief local time period (19 MLT), possibly due to negative stellar magnetic field B$_y$ (-5.25 nT). In the southern hemisphere, the polar cap expands to a maximum colatitude of 43.35° towards dusk during positive B$_y$. This agrees with antisymmetric dawn-dusk polar cap distributions between the two hemispheres \citep{Lukianova2011}. However, these shifts are lower than the equatorward shift of the open-close boundary (colatitude 54°) during a CME event from a young Sun of near similar age ($\sim$ 700 Myr) \citep{Airapetian2016}, indicating the intensity of the effects of a gradual CIR event was less than that of a CME event.  

Figure 9a shows one representative case of higher integrated energy fluxes from electron precipitation in the northern hemisphere from SWMF (RCM) simulation during the high dynamic pressure period. Electron precipitation is obtained from RCM by assuming 30\% of electrons are in the loss cone. Higher energy fluxes are localized and appear mainly near the dawn and dusk periods. The median of the average energy of these precipitating electrons is 1-6 keV, with a few average energies reaching up to $\sim$ 300-400 keV, especially during the high dynamic pressure or proton density periods, throughout the entire simulation duration. These higher energy electron precipitations can penetrate into altitudes below 70 km at mid and high latitudes of the Earth \citep{Mironova2015}. These higher integrated energy fluxes (Figure 9a) are associated with high FAC (maximum: 45.63 µA/m\textsuperscript{2}) (Figure 9b) and high height-integrated ionospheric Hall and Pedersen conductances at the same or near time instances (Figures 9c-d). The high precipitating electron flux, FAC, and ionospheric conductances occur at relatively lower latitudes. This may suggest the equatorward movement of the auroral oval region. 

\begin{figure*}[htbp]
\centering
\includegraphics[width=\textwidth,height=0.9\textheight,keepaspectratio]{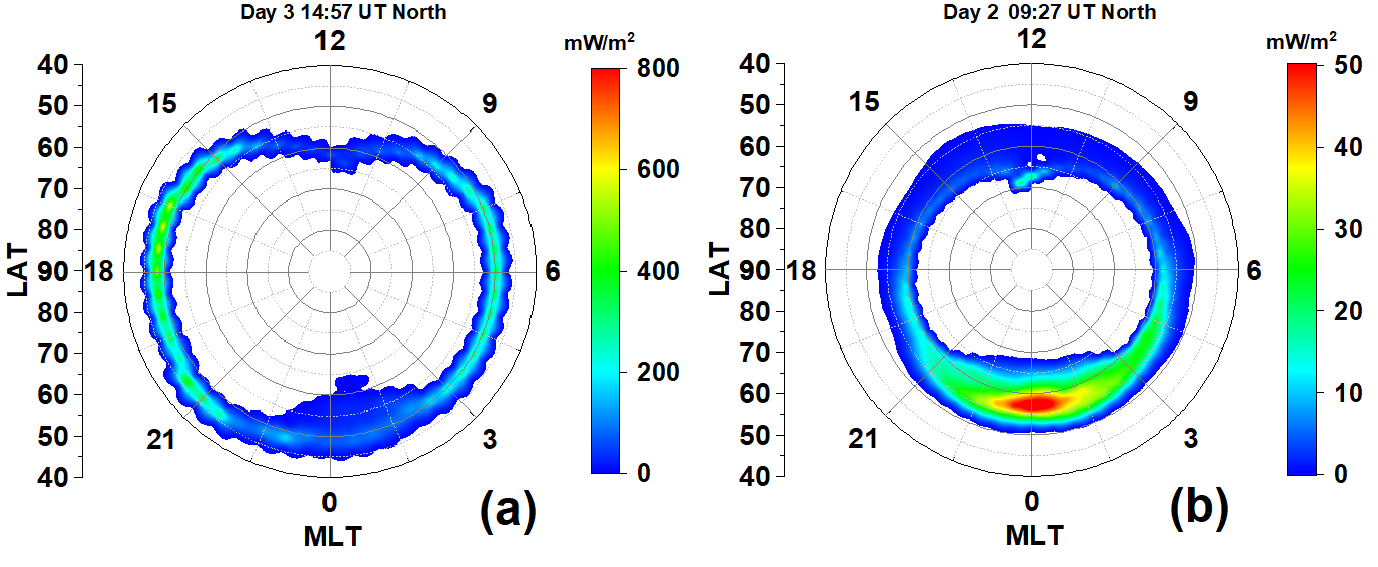}
\caption{Cases of northern hemispheric integrated energy fluxes from precipitating electrons during (a) high proton density period, and (b) high stellar wind speed period
\label{fig:general}}
\end{figure*}

During the entire simulation period, stellar wind speed varies between 691.86 and 1100.90 km/sec. Proton density varies between 122.75 and 1844.31/cm$^{3}$. We have conducted a comparative study of auroral electron precipitation during a high proton density and stellar wind speed period from the northern hemisphere, the two components of the wind dynamic pressure (shown in Figures 10a-b, respectively). The proton density at 14:57 UT, Day 3 was higher (1844.31/cm$^{3}$). Corresponding stellar wind speed was near smaller values inside the total period of simulation (746.14 km/sec). The stellar wind speed at 9:27 UT, Day 2 was 1100.59 km/sec. Corresponding proton density was lower (216.77/cm$^{3}$). We have found that the energy fluxes during the high proton density period are higher than those from the high stellar wind speed period. The electron precipitation is more equatorward during the higher proton density period. During the high proton density period, higher energy fluxes are localized and are seen mainly around dawn and dusk sectors. In contrast, higher energy fluxes during high stellar wind speed periods are seen near midnight than during dawn-dusk periods on several occasions.

It should be noted that the variation of stellar wind speed is $\sim$ 1.5 times for the two cases shown in Figure 10. However, the proton density ratio is $\sim$ 8.5 times. Moreover, as shown earlier, the stellar wind dynamic pressure is more correlated with proton density than stellar wind speed. Here, our simulation also suggests that the wind density plays dominant role in the electron precipitation compared to the wind speed. 

\begin{figure*}[htbp]
\centering
\includegraphics[width=\textwidth,height=0.9\textheight,keepaspectratio]{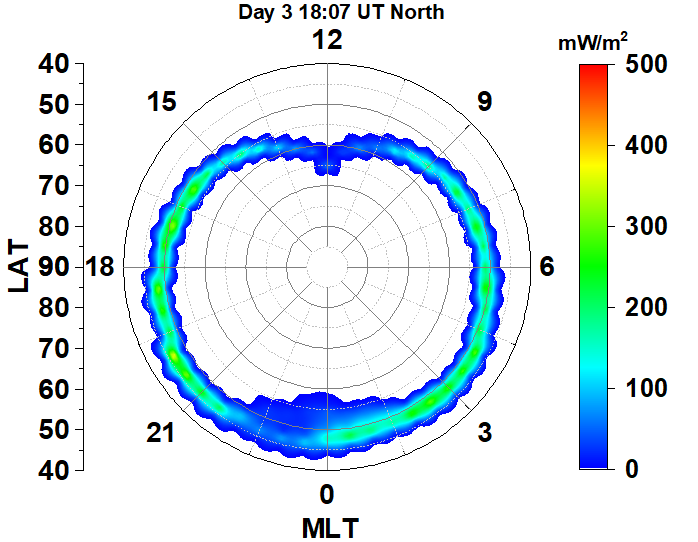}
\caption{Integrated energy flux from SWMF electron precipitation in the northern hemisphere during positive stellar magnetic field B$_z$ and low B$_y$ period with high proton density, dynamic pressure, and stellar wind speed
\label{fig:general}}
\end{figure*}

In the present SWMF+RCM simulation, there is a subset of time when the stellar magnetic field B$_z$ is northward and B$_y$ is very low ($|B_y| \leq 2$) during 17:59-23:32 UT, Day 3 (time span: over five and a half hours). The stellar wind speed is $\sim$ 700 km/sec, alongside varying proton density ($\sim$ 465-1108 /cm\textsuperscript{3}), and dynamic pressure ($\sim$ 470-1074 nPa) during this period. Therefore, we have used this subset of the total simulation period to observe the impacts of viscous interactions between stellar wind and the Earth's magnetosphere on the precipitating electron integrated energy flux, in the absence of significant IMF-driven reconnection. We have shown such a representative case during 18:07 UT, Day 3, when dynamic pressure (1033.1 nPa) and proton density (1071.86 /cm\textsuperscript{3}) values are high (Figure 11). The stellar wind speed is 694.20 km/sec. Therefore, the electron precipitation during this period may mostly be driven by the viscous interaction. The maximum integrated energy flux is 548 mW/m\textsuperscript{2} at this time instant. This is a relatively lower integrated energy flux than those shown in Figures 9a and 10a (maximum $>$ 800 mW/m\textsuperscript{2}, in the presence of a higher stellar magnetic field). This indicates that the overall electron precipitation is reduced during a low geoeffective interplanetary electric field period or active Dungey reconnection period. However, the number (548 mW/m\textsuperscript{2}) still indicates a significant and comparable amount of precipitation (much higher than the case shown in Figure 10b for lower proton density). Moreover, the electron precipitation shows MLT-wise asymmetries in the absence of the stellar magnetic field B$_y$ and during northward B$_z$. For example, precipitation enhances around 15-22 MLT and reduces around midnight (Figure 11). This may indicate the formation of partial ring currents due to intense viscous interactions. Therefore, this suggests the contribution of viscous interaction between stellar wind and the Earth's magnetosphere (by higher stellar proton density, dynamic pressure, and stellar wind speed) to the electron precipitation. The viscous interaction causes plasma injection into the ring current, and subsequently, the plasma gets scattered and precipitates into the loss cone. Overall, this roughly indicates that both Dungey reconnection and viscous interaction contribute to the electron precipitation in the entire simulation period. These results also indicate that the early Earth might have encountered these stronger effects more frequently than the present day. 

Therefore, the CIR storm simulated in this paper is mainly a pressure-driven event as the stellar wind dynamic pressure (and the proton density) are the major drivers, followed by the contributions from the stellar magnetic field B$_y$, stellar wind speed, and also possibly the smaller magnitude stellar magnetic field B$_z$. The magnetospheric and ionospheric responses to the event, such as ionospheric Joule heating, electron precipitation, FAC, and ionospheric conductances, are interrelated. Region 1 and Region 2 FAC are affected by the solar wind dynamic pressure, as shown by \cite{Iijima1982, Nakano2009}, respectively. This is due to the dynamic pressure driven compression of the Earth's magnetic field and increased dayside solar wind-magnetosphere coupling \citep{Cheng2021}. This can explain higher FAC during the higher dynamic pressure period (Figure 9b). Higher FAC causes the parallel electric field, accelerating the particles and increasing particle precipitation from the magnetosphere to the ionosphere \citep{Korth2014, Mukhopadhyay2022}. Moreover, previous studies suggested possible effects of solar wind dynamic pressure on whistler mode waves \citep{Yu2018} and Electromagnetic Ion Cyclotron (EMIC) waves \citep{Saikin2016}. These increase wave-particle interactions for electrons of varying energies. All these explain the higher precipitating electron integrated energy flux during the higher dynamic pressure period (Figure 9a). The increase in auroral particle precipitation during high dynamic pressure periods and both during northern and southern IMF B$_z$ was suggested by \cite{Craven1986, Li2003, Boudouridis2003}. Higher particle precipitation increases ionospheric conductances. Therefore, this and higher CPCP also explain higher atmospheric Joule heating (Figure 7a). \cite{Palmroth2004} also found correlations between dynamic pressure, FAC, and ionospheric Joule heating even during northward IMF conditions. Higher dynamic pressure compresses the Earth's magnetosphere. Therefore, the ionospheric footprints of the magnetic field lines also shift towards low latitudes. This also ensures equatorward movements of the ionospheric footprint of FAC \citep{Cheng2021}. Moreover, due to these shifts, the polar cap also expands as the open-close boundary also shifts to the lower latitudes. These may explain the equatorward shift of FAC (Figure 9b) and polar cap boundary (Figure 8) during the higher dynamic pressure period in the present simulation. These discussions may also explain the high integrated energy flux of auroral particle precipitation at lower latitudes mainly during high dynamic pressure and high proton density periods (Figures 9a-b, 10a). Moreover, the B$_y$ component stellar magnetic field and stellar wind speed are also higher in the present simulation. This supports stronger coupling between stellar wind and the Earth's magnetic field through the geoeffective interplanetary electric field, contributing to the equatorward expansion of the polar cap boundary and equatorward shifting of auroral precipitation.

We have already mentioned that proton density is more correlated with dynamic pressure in the present simulation than the stellar wind speed. Moreover, the proton density values in the present simulation are nearly 18 times higher than the typical values seen in recent days. Therefore, it also has dominant effects on electron precipitations than stellar wind speed during the CIR events from the young Sun ($\sim$ 600 Myr), as seen from Figure 10. We have seen good correlations between dynamic pressure and dayside magnetopause standoff distance, and hemispherically integrated Joule heating (Figures 2 and 7b, respectively). Therefore, these must show higher correlations with proton density than stellar wind speed. We have also performed calculations of Pearson's r coefficients regarding this. The correlation coefficients for dayside magnetopause standoff distance with dynamic pressure, proton density, and stellar wind speed are 0.89, 0.82, and 0.20, respectively. The correlation coefficients for northern hemispherically integrated Joule heating are 0.77, 0.59, and 0.27, respectively. The same for the southern hemisphere are 0.75, 0.67, and 0.07, respectively. Although we have seen higher equatorward expansion of the polar cap boundary during high dynamic pressure periods, the correlation of polar cap expansion with dynamic pressure is moderate. We have observed the correlation between the maximum equatorward polar cap expansion colatitude (combining all MLTs) with dynamic pressure, proton density, and stellar wind speed. The values are 0.29, 0.17, and 0.10, respectively. However, maximum equatorward polar cap expansion correlates better with stellar magnetic field B$_y$ (0.48). This indicates the effects of stellar magnetic field B$_y$ asymmetry on the polar cap boundaries. Moreover, this also indicates the dominant contribution of stellar magnetic field B$_y$ on the geoeffective interplanetary electric field over stellar magnetic field B$_z$, supporting the equatorward polar cap expansion in the present simulation.

\section{Discussions and Conclusions} \label{sec:highlight}

We have presented for the first time the numerical simulation of the effects of a CIR event from a young Sun ($\kappa^1$ Ceti, $\sim$ 600 Myr) on the magnetosphere and ionosphere of an early Earth. The simulation draws several important conclusions concerning the physical and chemical state of its upper atmosphere. The stellar inputs show that the proton density increases more than the stellar wind speed, contributing more to the dynamic pressure variations in this CIR event (higher correlation). Proton density values observed in this simulation is $\sim$ 18 times higher than those recorded during the disturbed conditions from the present Earth. However, the peak stellar wind speed from our simulation is inside the range of present-day disturbed condition values. The SWMF simulation shows higher dayside magnetopause compression than that during the May 10-12, 2024 super geomagnetic storm. The magnetopause standoff distance is correlated with dynamic pressure and proton density. Maximum negative SYMH is -703.90 nT, which is stronger than almost any values measured in severe geomagnetic storms since 1957-1958, but comparable with the March 1989 storm and weaker than the Carrington event. 
The high SYMH values may be explained by the large geoeffective interplanetary electric field, mainly due to stellar magnetic field B$_y$ and stellar wind speed. The negative peaks of SYMH are closely correlated with the stellar wind dynamic pressure. The first peak is higher, may be because of the presence of southward stellar magnetic field B$_z$. Apart from the ring current, the total SYMH may have contributions from other sources, such as magnetopause current, FAC, and cross-tail current, possibly due to extreme magnetopause compression by high stellar wind dynamic pressure. All these currents, including the ring current, are generally enhanced during higher dynamic pressure period. Therefore, Dst or SYMH is not an appropriate indicator for geomagnetic storm intensities during such an extreme condition.

Simulated CPCP is comparable to those observed during the October 29-31, 2003, geomagnetic storm. In addition to the dayside geoeffective interplanetary electric field, the CPCP is also significantly influenced by viscous interactions between stellar wind plasma and the Earth's magnetosphere. The interaction is mainly supported by the high dynamic pressure and proton density, and also associated with high stellar wind speed in order to ensure momentum transfer from the stellar wind to the Earth's magnetosphere. The simulated hemispheric Joule heating rates are comparable to the historical total global powers from NO and CO$_2$ radiative cooling rates observed during geomagnetically disturbed periods. The frequency of exposure of the early Earth's magnetosphere to the geomagnetic storms via CIRs is over two orders of magnitude greater than those observed today due to the shorter rotation period of the ${\kappa}^1$ Ceti, the proxy of the young Sun. This also contributes to the stronger effects of these phenomena during the periods of the young Sun compared to the present day. The polar cap boundary expands equatorward with the increase in dynamic pressure, especially at the nightside. The polar cap boundary also shows the expected asymmetries arising from the B$_y$ component of the stellar magnetic field. 
The equatorward expansion of the polar cap boundary is less correlated with dynamic pressure. However, it shows better correlations with stellar magnetic field B$_y$. This may be because the stellar magnetic field B$_y$ contributes to the geoeffective interplanetary electric field. The values of B$_y$ are higher than B$_z$ in our simulation, making this a more dominant contributor to the geoeffective interplanetary electric field. Higher correlations with stellar magnetic field B$_y$ may also be explained by the role of B$_y$ asymmetries on the polar cap expansions. As the dynamic pressure and proton density are more correlated than the stellar wind speed, hemispherically integrated Joule heating and precipitating electron fluxes to the ionosphere also increase with greater correlations with the dynamic pressure and proton density, compared to the stellar wind speed. The precipitation extends equatorward also mainly during higher proton density periods. Moreover, the precipitation fluxes exhibit localized enhancements near the dawn-dusk side during higher proton density periods. However, the precipitating particle fluxes maximize near midnight during higher stellar wind speed periods in most cases. Higher precipitated electron fluxes during high dynamic pressure periods are also associated with higher FAC and height-integrated ionospheric Hall and Pedersen conductances. However, the dayside magnetopause standoff distance is farther from the Earth, and the equatorward expansion of the polar cap is lower during the CIR event (in our simulation) than those observed from the simulation of a CME event from the young Sun of similar age \citep{Airapetian2016}. Overall, the results show an extreme dynamic pressure or proton density driven storm event, which is also supported by stellar wind speed, stellar magnetic field B$_y$, and a smaller B$_z$ components. The dayside reconnection is mainly influenced by the stellar magnetic field B$_y$ component. The extreme compression of the magnetosphere may also affect the ring current and the Dst. The magnetospheric and ionospheric responses, such as CPCP or auroral electron precipitation, are also highly influenced by viscous interactions between the stellar wind and the Earth's magnetosphere.

However, the models may not work satisfactorily during such an extreme event (with a highly compressed magnetosphere), even though SWMF is proven to be useful in such circumstances \citep{Airapetian2016}. For example, as mentioned above, the Dst obtained from SWMF may be influenced with currents from different sources apart from the ring current. These could be magnetopause current or FAC \citep{Blake2021}. Different current sources, such as FAC, magnetospheric and ionospheric currents are responsible for the distortion of the Earth's magnetic field during an extreme event \citep{Thomas2024}. This is mainly because of the extremely high dynamic pressure, which has compressed the magnetosphere. Though well-separated from the boundary effects, the RCM outputs at the closed field line region can still be influenced by the stellar wind plasma injection due to the heavily compressed structure of the magnetosphere. 
The simulation is conducted using the present-day Earth dipole configuration and magnetic strength. The simulation needs to be done with early Earth configurations, such as the early Earth dipole magnetic field strength and orientation, for more realistic scenarios. The solar F10.7 index is provided according to recent time values. It must be adjusted to the early Sun-Earth conditions, especially for the accurate dayside ionospheric conductance due to solar illumination. The accurate F10.7 should be fitted in the relations given by \cite{Moen1993}. The thermospheric compositions during the early Earth were different and this needs to be properly considered. In the future study, we will precisely implement the magnetosphere-ionosphere-thermosphere coupling during the early Earth. This will be done by coupling the SWMF+RCM models with ionosphere-thermosphere models in the next stage. The overall setup will provide realistic information regarding atmospheric heating, atmospheric escape, global circulation patterns, and neutral wind interactions under the early Earth conditions. Another drawback is that SWMF uses an ionospheric conductance model, which was derived using the empirical AMIE method \citep{Ahn1998} during January 1997 (mostly geomagnetically quiet period) \citep{Ridley2004, Welling2019, Mukhopadhyay2020}. This relation may not be suitable for extreme stellar wind conditions. Therefore, there is always some uncertainty in the overall SWMF model performance during this extreme event. Moreover, ionospheric conductances must also be estimated separately for different types of precipitations for better accuracy, especially during extreme solar wind conditions. Conductance estimated using the first principle could be suitable if it is incorporated with altitude-dependent ionospheric and thermospheric electrodynamics. However, this is a very challenging process \citep{Mukhopadhyay2022}. The version of RCM used to couple with SWMF uses a simplistic time constant for the electron scattering at the loss cone \citep{Chen2019}. This could yield differences from the real-time scenario. Moreover, RCM assumes an isotropic pitch angle distribution at the plasma sheet. This is an oversimplification 
for the ring current particle distributions \citep{Fok2001}. As already mentioned, the model has to be simulated in CCMC low resolution mode to avoid the simulation being terminated abruptly. Higher resolution might have given a better picture of the stellar wind-magnetosphere interaction. However, efforts are underway to obtain higher resolution runs alongside other possible improvements mentioned above for the better visualization of these extreme conditions.

These observations indicate that frequent CIR events during the young Sun epoch could trigger various geomagnetic and atmospheric changes. For example, the higher compression of the magnetopause standoff distance and subsequent substantial expansion in the polar cap region result in a larger area for ion outflow through open magnetic field lines. Enhanced Joule heating rates heat and expand the upper atmosphere, increase the ion scale height \citep{Cohen2014, Cohen2018}, and accelerate the ion escape \citep{Airapetian2017}. Finally, the higher electron precipitation drives the enhanced ion escape via increased local ionization rate of O+ and N+ ions, collisional heating, and the ambipolar electric field at the Earth's exosphere. This also increases the scale height and drives ion outflow. These effects are elaborated in \cite{Strangeway2005, Airapetian2017, Airapetian2020}. All phenomena observed in this study, especially high-energy electron precipitation ($\sim$ 300-400 keV), could also affect the atmospheric chemistry below 70 km \citep{Mironova2015, Airapetian2020} and may be important to the habitability factors of an Earth-like planet around an active star. The role of frequent CME and CIR events in the evolution of early Venusian and Martian atmospheres will be addressed in a forthcoming study. This includes addressing how Mars lost its magnetic field and subsequently most of its atmosphere, greatly reducing habitability conditions.

Our study has important implications for understanding the state of Earth's upper atmosphere and its potential habitability during the early to mid-Hadean eon (first 600 Myr), a period increasingly recognized as critical for the origin of life \citep{Wilde2001, Pearce2018}. Recent observational studies suggest that young solar analogs within the first 100–200 Myr rotate approximately three times faster than ${\kappa}^1$ Ceti and drive stronger stellar winds, resulting in higher CIR driven dynamic pressure and that the early Earth may spend over 30$\%$ of the time in the compressed regions \citep{Namekata2024a}. Also, the Sun produced frequent and energetic CMEs that presumably generated high-fluence hard-spectra solar energetic particle events. These particles precipitated in the low atmosphere of early Earth and initiated efficient generation of biologically important molecules for the origin of life on Earth, including amino acids and carboxylic acids \citep{Airapetian2016, Hu2022, Kobayashi2023}. These processes and associated prebiotic chemistry in the planetary atmosphere would also be a plausible scenario for the rise of habitability on young rocky exoplanets around magnetically active solar-like stars. Overall, similar simulations of the state of the upper atmosphere, as presented in this paper and other studies, during different phases of the early Earth evolution, are extremely important to understand the planetary habitability. The knowledge gained from these simulations would be critical for searching for life on rocky exoplanets with James Webb Space Telescope (JWST) and upcoming LIFE and Habitable World Observatory (HWO).

\begin{acknowledgments}
We would like to thank Professor Daniel Welling, the referee of the paper, for constructive suggestions that greatly improved this paper. Dibyendu Sur is supported through NASA Cooperative Agreement 80NSSC21M0180 under the Partnership for Heliophysics and Space Environment Research (PHaSER). Vladimir S. Airapetian acknowledges support from the NASA/GSFC Sellers Exoplanet Environments Collaboration (SEEC), which is funded by the NASA Planetary Science Division’s Internal Scientist Funding Model (ISFM), NASA's Astrophysics Theory Program grant \#80NSSC24K0776, HST-GO-15825.004-A and NICER Cycle 2 programs. The model simulation results were obtained from CCMC at NASA Goddard Space Flight Center (GSFC), via their public Runs on Request system (website: ccmc.gsfc.nasa.gov). The CCMC operates as a collaborative effort among multiple agencies, including NASA, AFMC, AFOSR, AFRL, AFWA, NOAA, NSF, and ONR. The authors thank Dr. Mei-Ching Fok, NASA GSFC, Dr. Lutz Rastaetter, NASA GSFC and CCMC, Dr. Darren De Zeeuw, the Catholic University of America and NASA GSFC, and Dr. David Sibeck, NASA GSFC, for their suggestions and immense help towards the paper. BATS-R-US and overall SWMF codes were developed by the Center for Space Environment Modeling (CSEM) at the University of Michigan. All these tools are available via the University of Michigan for licensed users. These can also be accessed through the CCMC. The Dst, SYMH, and different solar wind parameter data during different geomagnetically disturbed periods are obtained from NASA/GSFC Space Physics Data Facility's (SPDF) OMNIWeb at the website omniweb.gsfc.nasa.gov. Dst and SYMH data are computed at The World Data Center for Geomagnetism (WDC), University of Kyoto, at the website wdc.kugi.kyoto-u.ac.jp/aeasy/ and all its contributing data providers.  
\end{acknowledgments}

\bibliographystyle{aasjournal}
\bibliography{ms}

\begin{thebibliography}{}
\expandafter\ifx\csname natexlab\endcsname\relax\def\natexlab#1{#1}\fi
\providecommand{\url}[1]{\href{#1}{#1}}
\providecommand{\dodoi}[1]{doi:~\href{http://doi.org/#1}{\nolinkurl{#1}}}
\providecommand{\doeprint}[1]{\href{http://ascl.net/#1}{\nolinkurl{http://ascl.net/#1}}}
\providecommand{\doarXiv}[1]{\href{https://arxiv.org/abs/#1}{\nolinkurl{https://arxiv.org/abs/#1}}}

\bibitem[{Ahn {et~al.}(1998)Ahn, Richmond, Kamide, Kroehl, Emery, de~la Beaujardi\`{e}re, \& Akasofu}]{Ahn1998}
Ahn, B.-H., Richmond, A.~D., Kamide, Y., {et~al.} 1998, Journal of Geophysical Research, 103, 14769, \dodoi{10.1029/97JA03088}

\bibitem[{Airapetian {et~al.}(2020)Airapetian, Barnes, Cohen, Collinson, Danchi, {et~al.}}]{Airapetian2020}
Airapetian, V.~S., Barnes, R., Cohen, O., {et~al.} 2020, International Journal of Astrobiology, 19, 136

\bibitem[{Airapetian {et~al.}(2016)Airapetian, Glocer, Gronoff, Hébrard, \& Danchi}]{Airapetian2016}
Airapetian, V.~S., Glocer, A., Gronoff, G., Hébrard, E., \& Danchi, W. 2016, Nature Geoscience, 9, 452

\bibitem[{Airapetian {et~al.}(2017)Airapetian, Glocer, Khazanov, Loyd, France, {et~al.}}]{Airapetian2017}
Airapetian, V.~S., Glocer, A., Khazanov, G.~V., {et~al.} 2017, The Astrophysical Journal Letters, 836, L3

\bibitem[{Airapetian {et~al.}(2021)Airapetian, Jin, Lüftinger, Saikia, {et~al.}}]{Airapetian2021}
Airapetian, V.~S., Jin, M., Lüftinger, T., Saikia, S.~B., {et~al.} 2021, The Astrophysical Journal, 916, 96

\bibitem[{Akasofu(1991)}]{Akasofu1991}
Akasofu, S.-I. 1991, in Geophysical Monograph 26, ed. C.-I. Meng, M.~J. Rycroft, \& L.~A. Frank (Cambridge: Cambridge University Press), 1--12

\bibitem[{Axford(1964)}]{Axford1964}
Axford, W.~I. 1964, Planetary and Space Science, 12, 45, \dodoi{10.1016/0032-0633(64)90067-4}

\bibitem[{Axford \& Hines(1961)}]{AxfordHines1961}
Axford, W.~I., \& Hines, C.~O. 1961, Canadian Journal of Physics, 39, 1433, \dodoi{10.1139/p61-172}

\bibitem[{Blake {et~al.}(2021)Blake, Pulkkinen, Schuck, Glocer, Oliveira, Welling, {et~al.}}]{Blake2021}
Blake, S.~P., Pulkkinen, A., Schuck, P.~W., {et~al.} 2021, Space Weather, 19, e2020SW002585, \dodoi{10.1029/2020SW002585}

\bibitem[{Borovsky \& Denton(2006)}]{Borovsky2006}
Borovsky, J.~E., \& Denton, M.~H. 2006, Journal of Geophysical Research, 111

\bibitem[{Boudouridis {et~al.}(2007)Boudouridis, Lyons, Zesta, \& Ruohoniemi}]{Boudouridis2007}
Boudouridis, A., Lyons, L.~R., Zesta, E., \& Ruohoniemi, J.~M. 2007, Journal of Geophysical Research: Space Physics, 112, A06201, \dodoi{10.1029/2006JA012141}

\bibitem[{Boudouridis {et~al.}(2003)Boudouridis, Zesta, Lyons, Anderson, \& Lummerzheim}]{Boudouridis2003}
Boudouridis, A., Zesta, E., Lyons, L.~R., Anderson, P.~C., \& Lummerzheim, D. 2003, Journal of Geophysical Research: Space Physics, 108, 8012, \dodoi{10.1029/2002JA009373}

\bibitem[{Chen {et~al.}(2019)Chen, Lemon, Hecht, Sazykin, Wolf, Boyd, \& Valek}]{Chen2019}
Chen, M.~W., Lemon, C.~L., Hecht, J., {et~al.} 2019, Journal of Geophysical Research: Space Physics, 124, 4194, \dodoi{https://doi.org/10.1029/2019JA026545}

\bibitem[{Cheng {et~al.}(2014)Cheng, Shi, Dunlop, \& Liu}]{Cheng2014}
Cheng, Z.~W., Shi, J.~K., Dunlop, M., \& Liu, Z.~X. 2014, Journal of Atmospheric and Solar-Terrestrial Physics, 115--116, 52, \dodoi{10.1016/j.jastp.2013.11.003}

\bibitem[{Cheng {et~al.}(2021)Cheng, Shi, Torkar, Lu, Dunlop, Carr, {et~al.}}]{Cheng2021}
Cheng, Z.~W., Shi, J.~K., Torkar, K., {et~al.} 2021, Journal of Geophysical Research: Space Physics, 126, e2021JA029785, \dodoi{10.1029/2021JA029785}

\bibitem[{Cohen {et~al.}(2014)Cohen, Drake, Glocer, Garraffo, Poppenhaeger, {et~al.}}]{Cohen2014}
Cohen, O., Drake, J.~J., Glocer, A., {et~al.} 2014, The Astrophysical Journal, 790, 57

\bibitem[{Cohen {et~al.}(2018)Cohen, Glocer, Garraffo, Drake, \& Bell}]{Cohen2018}
Cohen, O., Glocer, A., Garraffo, C., Drake, J.~J., \& Bell, J.~M. 2018, The Astrophysical Journal Letters, 856, L11

\bibitem[{Craven {et~al.}(1986)Craven, Frank, Russell, Smith, \& Lepping}]{Craven1986}
Craven, J.~D., Frank, L.~A., Russell, C.~T., Smith, E.~J., \& Lepping, R.~P. 1986, in Solar Wind--Magnetosphere Coupling, ed. Y.~Kamide \& J.~A. Slavin (Tokyo: Terra Scientific), 367--380

\bibitem[{Da~Silva {et~al.}(2025)Da~Silva, Shi, Alves, Resende, Vieira, Costa, Marchezi, Agapitov, Sibeck, dos Santos, Andrioli, Jauer, Deggeroni, do~Carmo, Nyassor, Chen, Ayorinde, Ferreira, Moro, Wang, Li, \& Liu}]{DaSilva2025}
Da~Silva, L.~A., Shi, J., Alves, L.~R., {et~al.} 2025, Frontiers in Astronomy and Space Sciences, 12, 1550635, \dodoi{10.3389/fspas.2025.1550635}

\bibitem[{De~Zeeuw {et~al.}(2004)De~Zeeuw, Sazykin, Wolf, Gombosi, Ridley, \& Tóth}]{DeZeeuw2004}
De~Zeeuw, D.~L., Sazykin, S., Wolf, R.~A., {et~al.} 2004, Journal of Geophysical Research, 109, A12219

\bibitem[{Dessler \& Parker(1959)}]{Dessler1959}
Dessler, A.~J., \& Parker, E.~N. 1959, Journal of Geophysical Research, 64, 2239, \dodoi{10.1029/JZ064i012p02239}

\bibitem[{Dmitriev {et~al.}(2014)Dmitriev, Suvorova, Chao, Wang, Rastaetter, Panasyuk, Lazutin, Kovtyukh, Veselovsky, \& Myagkova}]{Dmitriev2014}
Dmitriev, A.~V., Suvorova, A.~V., Chao, J.-K., {et~al.} 2014, Journal of Geophysical Research: Space Physics, 119, 877, \dodoi{10.1002/2013JA019534}

\bibitem[{Du {et~al.}(2012)Du, Tsurutani, \& Sun}]{DU201255}
Du, A., Tsurutani, B., \& Sun, W. 2012, Planetary and Space Science, 71, 55, \dodoi{https://doi.org/10.1016/j.pss.2012.07.002}

\bibitem[{Du {et~al.}(2008)Du, Tsurutani, \& Sun}]{Du2008}
Du, A.~M., Tsurutani, B.~T., \& Sun, W. 2008, Journal of Geophysical Research: Space Physics, 113, A10214, \dodoi{10.1029/2008JA013284}

\bibitem[{Dungey(1961)}]{Dungey1961}
Dungey, J.~W. 1961, Physical Review Letters, 6, 47

\bibitem[{Elphinstone {et~al.}(1993)Elphinstone, Hearn, Murphree, Cogger, Johnson, \& Vo}]{Elphinstone1993}
Elphinstone, R.~D., Hearn, D.~J., Murphree, J.~S., {et~al.} 1993, in Geophysical Monograph Series, Vol.~80, Auroral Plasma Dynamics, ed. R.~L. Lysak (American Geophysical Union), \dodoi{10.1029/GM080p0031}

\bibitem[{Fedder {et~al.}(1991)Fedder, Mobarry, \& Lyon}]{Fedder1991}
Fedder, J.~A., Mobarry, C.~M., \& Lyon, J.~G. 1991, Geophysical Research Letters, 18, 1047, \dodoi{10.1029/90GL02722}

\bibitem[{Fejer {et~al.}(1990)Fejer, Spiro, Wolf, \& Foster}]{Fejer1990}
Fejer, B.~G., Spiro, R.~W., Wolf, R.~A., \& Foster, J.~C. 1990, Annales Geophysicae, 8, 441

\bibitem[{Fok {et~al.}(2001)Fok, Wolf, Spiro, \& Moore}]{Fok2001}
Fok, M.-C., Wolf, R.~A., Spiro, R.~W., \& Moore, T.~E. 2001, Journal of Geophysical Research: Space Physics, 106, 8417, \dodoi{https://doi.org/10.1029/2000JA000235}

\bibitem[{Fu {et~al.}(2025)Fu, Fu, Zhang, Yu, \& Cao}]{Fu2025}
Fu, W.~D., Fu, H.~S., Zhang, W.~Z., Yu, Y., \& Cao, J.~B. 2025, Geophysical Research Letters, 52, e2024GL114040

\bibitem[{Fuselier {et~al.}(2024)Fuselier, Petrinec, Reiff, {et~al.}}]{Fuselier2024}
Fuselier, S.~A., Petrinec, S.~M., Reiff, P.~H., {et~al.} 2024, Space Science Reviews, 220, 34, \dodoi{10.1007/s11214-024-01067-0}

\bibitem[{Garcia-Sage {et~al.}(2017)Garcia-Sage, Glocer, Drake, Gronoff, \& Cohen}]{GarciaSage2017}
Garcia-Sage, K., Glocer, A., Drake, J.~J., Gronoff, G., \& Cohen, O. 2017, The Astrophysical Journal Letters, 844, L13, \dodoi{10.3847/2041-8213/aa7eca}

\bibitem[{Glocer {et~al.}(2009)Glocer, T{\'o}th, Fok, Gombosi, \& Liemohn}]{Glocer2009}
Glocer, A., T{\'o}th, G., Fok, M., Gombosi, T., \& Liemohn, M. 2009, Journal of Atmospheric and Solar-Terrestrial Physics, 71, 1653

\bibitem[{Gombosi {et~al.}(2021)Gombosi, Chen, Glocer, Huang, Jia, \& et~al.}]{Gombosi2021}
Gombosi, T.~I., Chen, Y., Glocer, A., {et~al.} 2021, Journal of Space Weather and Space Climate, 11, 42

\bibitem[{Gombosi {et~al.}(2004)Gombosi, Powell, De~Zeeuw, Clauer, Hansen, Manchester, Ridley, Roussev, Sokolov, Stout, \& T{\'o}th}]{Gombosi2004}
Gombosi, T.~I., Powell, K.~G., De~Zeeuw, D.~L., {et~al.} 2004, Computing in Science \& Engineering, 6, 14, \dodoi{10.1109/MCISE.2004.1267603}

\bibitem[{Gonzalez \& Mozer(1974)}]{Gonzalez1974}
Gonzalez, W.~D., \& Mozer, F.~S. 1974, Journal of Geophysical Research, 79, 4186, \dodoi{10.1029/JA079i028p04186}

\bibitem[{Gonzalez {et~al.}(1999)Gonzalez, Tsurutani, \& Cluá~de Gonzalez}]{Gonzalez1999}
Gonzalez, W.~D., Tsurutani, B.~T., \& Cluá~de Gonzalez, A.~L. 1999, Space Science Reviews, 88, 529

\bibitem[{Harel {et~al.}(1981)Harel, Wolf, Reiff, Spiro, Burke, Rich, \& Smiddy}]{Harel1981}
Harel, M., Wolf, R.~A., Reiff, P.~H., {et~al.} 1981, Journal of Geophysical Research, 86, 2217

\bibitem[{Harnett {et~al.}(2008)Harnett, Winglee, Stickle, \& Lu}]{Harnett2008}
Harnett, E.~M., Winglee, R.~M., Stickle, A., \& Lu, G. 2008, Journal of Geophysical Research, 113, A06209

\bibitem[{Hu {et~al.}(2022)Hu, Airapetian, Li, Zank, \& Jin}]{Hu2022}
Hu, J., Airapetian, V.~S., Li, G., Zank, G., \& Jin, M. 2022, Science Advances, 8, eabi9743

\bibitem[{Huang {et~al.}(2007)Huang, Burke, \& Lin}]{Huang2007}
Huang, C.~Y., Burke, W.~J., \& Lin, C.~S. 2007, Journal of Atmospheric and Solar-Terrestrial Physics, 69, 101

\bibitem[{Hudson {et~al.}(2014)Hudson, Baker, Goldstein, Kress, Paral, Toffoletto, \& Wiltberger}]{Hudson2014}
Hudson, M.~K., Baker, D.~N., Goldstein, J., {et~al.} 2014, Geophysical Research Letters, 41, 1113, \dodoi{https://doi.org/10.1002/2014GL059222}

\bibitem[{Iijima \& Potemra(1982)}]{Iijima1982}
Iijima, T., \& Potemra, T.~A. 1982, Geophysical Research Letters, 9, 442, \dodoi{10.1029/GL009i004p00442}

\bibitem[{Kan \& Lee(1979)}]{Kan1979}
Kan, J.~R., \& Lee, L.~C. 1979, Geophysical Research Letters, 6, 577, \dodoi{10.1029/GL006i007p00577}

\bibitem[{Kim {et~al.}(2020)Kim, Pogorelov, Arge, {et~al.}}]{Kim2020}
Kim, T.~K., Pogorelov, N.~V., Arge, C.~N., {et~al.} 2020, The Astrophysical Journal Supplement Series, 246

\bibitem[{Kobayashi {et~al.}(2023)Kobayashi, Ise, Aoki, Kinoshita, Naito, {et~al.}}]{Kobayashi2023}
Kobayashi, K., Ise, J.-i., Aoki, R., {et~al.} 2023, Life, 13, 1103

\bibitem[{Korth {et~al.}(2014)Korth, Zhang, Anderson, Sotirelis, \& Waters}]{Korth2014}
Korth, H., Zhang, Y., Anderson, B.~J., Sotirelis, T., \& Waters, C.~L. 2014, Journal of Geophysical Research: Space Physics, 119, 6715, \dodoi{10.1002/2014JA019961}

\bibitem[{Kuznetsova \& Laptukhov(2011)}]{Kuznetsova2011}
Kuznetsova, T.~V., \& Laptukhov, A.~I. 2011, Advances in Space Research, 47, 978, \dodoi{10.1016/j.asr.2010.11.022}

\bibitem[{Li {et~al.}(2003)Li, Baker, Elkington, Temerin, Reeves, Belian, Blake, Singer, Peria, \& Parks}]{Li2003}
Li, X., Baker, D.~N., Elkington, S., {et~al.} 2003, Journal of Atmospheric and Solar-Terrestrial Physics, 65, 233, \dodoi{10.1016/S1364-6826(02)00286-9}

\bibitem[{Liemohn {et~al.}(2018)Liemohn, Ganushkina, De~Zeeuw, Rastätter, Kuznetsova, Welling, \& et~al.}]{Liemohn2018}
Liemohn, M., Ganushkina, N.~Y., De~Zeeuw, D.~L., {et~al.} 2018, Space Weather, 16, 1583

\bibitem[{Liemohn(2003)}]{Liemohn2003}
Liemohn, M.~W. 2003, Journal of Geophysical Research, 108, 1251, \dodoi{10.1029/2003JA009839}

\bibitem[{Liemohn {et~al.}(2021)Liemohn, Shane, Azari, Petersen, Swiger, \& Mukhopadhyay}]{Liemohn2021}
Liemohn, M.~W., Shane, A.~D., Azari, A.~R., {et~al.} 2021, Journal of Atmospheric and Solar-Terrestrial Physics, 218, 105624, \dodoi{10.1016/j.jastp.2021.105624}

\bibitem[{Lukianova \& Kozlovsky(2011)}]{Lukianova2011}
Lukianova, R., \& Kozlovsky, A. 2011, Annales Geophysicae, 29, 1305

\bibitem[{Mironova {et~al.}(2015)Mironova, Aplin, Arnold, Bazilevskaya, Harrison, {et~al.}}]{Mironova2015}
Mironova, I.~A., Aplin, K.~L., Arnold, F., {et~al.} 2015, Space Science Reviews, 194, 1

\bibitem[{Mlynczak {et~al.}(2024)Mlynczak, Hunt, Nowak, Marshall, \& Mertens}]{Mlynczak2024}
Mlynczak, M.~G., Hunt, L.~A., Nowak, N., Marshall, B.~T., \& Mertens, C.~J. 2024, Geophysical Research Letters, 51, e2024GL110701

\bibitem[{Moen \& Brekke(1993)}]{Moen1993}
Moen, J., \& Brekke, A. 1993, Geophysical Research Letters, 20, 971, \dodoi{10.1029/92GL02109}

\bibitem[{Mukhopadhyay {et~al.}(2020)Mukhopadhyay, Welling, Liemohn, Ridley, Chakraborty, \& Anderson}]{Mukhopadhyay2020}
Mukhopadhyay, A., Welling, D.~T., Liemohn, M.~W., {et~al.} 2020, Space Weather, 18, e2020SW002551, \dodoi{10.1029/2020SW002551}

\bibitem[{Mukhopadhyay {et~al.}(2022)Mukhopadhyay, Welling, Liemohn, Ridley, Burleigh, Wu, Zou, Connor, Vandegriff, Dredger, \& T{\'o}th}]{Mukhopadhyay2022}
Mukhopadhyay, A., Welling, D., Liemohn, M., {et~al.} 2022, Journal of Geophysical Research: Space Physics, 127, e2022JA030323, \dodoi{10.1029/2022JA030323}

\bibitem[{Nakano {et~al.}(2009)Nakano, Ueno, Ohtani, \& Higuchi}]{Nakano2009}
Nakano, S., Ueno, G., Ohtani, S., \& Higuchi, T. 2009, Journal of Geophysical Research: Space Physics, 114, \dodoi{https://doi.org/10.1029/2008JA013674}

\bibitem[{Namekata {et~al.}(2024{\natexlab{a}})Namekata, Airapetian, Petit, {et~al.}}]{Namekata2024a}
Namekata, K., Airapetian, V.~S., Petit, P., {et~al.} 2024{\natexlab{a}}, The Astrophysical Journal, 961, 23

\bibitem[{Namekata {et~al.}(2024{\natexlab{b}})Namekata, Ikuta, Petit, Airapetian, Vidotto, Heinzel, Wollmann, Maehara, Notsu, \& Inoue}]{Namekata2024b}
Namekata, K., Ikuta, K., Petit, P., {et~al.} 2024{\natexlab{b}}, The Astrophysical Journal, 976, 255

\bibitem[{Newell {et~al.}(2007)Newell, Sotirelis, Liou, Meng, \& Rich}]{Newell2007}
Newell, P.~T., Sotirelis, T., Liou, K., Meng, C.-I., \& Rich, F.~J. 2007, Journal of Geophysical Research: Space Physics, 112, A01206, \dodoi{10.1029/2006JA012015}

\bibitem[{Ober {et~al.}(2006)Ober, Wilson, Maynard, Burke, \& Siebert}]{Ober2006}
Ober, D.~M., Wilson, G.~R., Maynard, N.~C., Burke, W.~J., \& Siebert, K.~D. 2006, Geophysical Research Letters, 33, \dodoi{https://doi.org/10.1029/2005GL024655}

\bibitem[{Palmroth {et~al.}(2004)Palmroth, Janhunen, Pulkkinen, \& Koskinen}]{Palmroth2004}
Palmroth, M., Janhunen, P., Pulkkinen, T.~I., \& Koskinen, H. E.~J. 2004, Annales Geophysicae, 22, 549, \dodoi{10.5194/angeo-22-549-2004}

\bibitem[{Pearce {et~al.}(2018)Pearce, Tupper, Pudritz, \& Higgs}]{Pearce2018}
Pearce, B. K.~D., Tupper, A.~S., Pudritz, R.~E., \& Higgs, P.~G. 2018, Astrobiology, 18, 343

\bibitem[{Petrinec {et~al.}(2016)Petrinec, Burch, Fuselier, Gomez, Lewis, Trattner, Ergun, Mauk, Pollock, Schiff, Strangeway, Russell, Phan, \& Young}]{Petrinec2016}
Petrinec, S.~M., Burch, J.~L., Fuselier, S.~A., {et~al.} 2016, Geophysical Research Letters, 43, 5997, \dodoi{10.1002/2016GL069626}

\bibitem[{Powell {et~al.}(1999)Powell, Roe, Linde, Gombosi, \& De~Zeeuw}]{Powell1999}
Powell, K.~G., Roe, P.~L., Linde, T.~J., Gombosi, T.~I., \& De~Zeeuw, D.~L. 1999, Journal of Computational Physics, 154, 284

\bibitem[{Priyadarshi {et~al.}(2021)Priyadarshi, Yang, \& Sun}]{Priyadarshi2021}
Priyadarshi, S., Yang, J., \& Sun, W. 2021, Frontiers in Astronomy and Space Sciences, 8, 691000

\bibitem[{Rastätter {et~al.}(2016)Rastätter, Shim, Kuznetsova, Kilcommons, Knipp, Codrescu, Fuller-Rowell, Emery, Weimer, Cosgrove, \& et~al.}]{Rastatter2016}
Rastätter, L., Shim, J.~S., Kuznetsova, M.~M., {et~al.} 2016, Space Weather, 14, 113

\bibitem[{Richmond \& Kamide(1988)}]{Richmond1988}
Richmond, A.~D., \& Kamide, Y. 1988, Journal of Geophysical Research, 93, 5741, \dodoi{10.1029/JA093iA06p05741}

\bibitem[{Ridley {et~al.}(2001)Ridley, De~Zeeuw, Gombosi, \& Powell}]{Ridley2001}
Ridley, A.~J., De~Zeeuw, D.~L., Gombosi, T.~I., \& Powell, K.~G. 2001, Journal of Geophysical Research, 106, 30067

\bibitem[{Ridley {et~al.}(2004)Ridley, Gombosi, \& DeZeeuw}]{Ridley2004}
Ridley, A.~J., Gombosi, T.~I., \& DeZeeuw, D.~L. 2004, Annales Geophysicae, 22, 567, \dodoi{10.5194/angeo-22-567-2004}

\bibitem[{Rosing(1999)}]{Rosing1999}
Rosing, M.~T. 1999, Science, 283, 674

\bibitem[{Rufenach {et~al.}(1992)Rufenach, McPherron, \& Schaper}]{Rufenach1992}
Rufenach, C.~L., McPherron, R.~L., \& Schaper, J. 1992, Journal of Geophysical Research, 97, 25, \dodoi{10.1029/91JA02135}

\bibitem[{Russell \& Huddleston(1997)}]{Russell1997}
Russell, C.~T., \& Huddleston, D.~E. 1997, Advances in Space Research, 20, 327, \dodoi{10.1016/S0273-1177(97)00681-9}

\bibitem[{Saikin {et~al.}(2016)Saikin, Zhang, Smith, Spence, Torbert, \& Kletzing}]{Saikin2016}
Saikin, A.~A., Zhang, J.-C., Smith, C.~W., {et~al.} 2016, Journal of Geophysical Research: Space Physics, 121, 4362, \dodoi{10.1002/2016JA022523}

\bibitem[{Sazykin(2000)}]{Sazykin2000}
Sazykin, S. 2000, Ph.d. thesis, Utah State University

\bibitem[{Sazykin {et~al.}(2002)Sazykin, Wolf, Spiro, Gombosi, De~Zeeuw, \& Thomsen}]{Sazykin2002}
Sazykin, S., Wolf, R.~A., Spiro, R.~W., {et~al.} 2002, Geophysical Research Letters, 29, 88

\bibitem[{Schrijver {et~al.}(2015)}]{Schrijver2015}
Schrijver, C.~J., {et~al.} 2015, Advances in Space Research, 55, 2745

\bibitem[{Sckopke(1966)}]{Sckopke1966}
Sckopke, N. 1966, Journal of Geophysical Research, 71, 3125, \dodoi{10.1029/JZ071i013p03125}

\bibitem[{Sibeck \& Lin(2014)}]{Sibeck2014}
Sibeck, D.~G., \& Lin, R.-Q. 2014, Journal of Geophysical Research: Space Physics, 119, 1028, \dodoi{10.1002/2013JA019471}

\bibitem[{Skoug {et~al.}(2004)Skoug, Gosling, Steinberg, McComas, Smith, {et~al.}}]{Skoug2004}
Skoug, R.~M., Gosling, J.~T., Steinberg, J.~T., {et~al.} 2004, Journal of Geophysical Research, 109, A09102

\bibitem[{Spiro {et~al.}(1981)Spiro, Harel, Wolf, \& Reiff}]{Spiro1981}
Spiro, R.~W., Harel, M., Wolf, R.~A., \& Reiff, P.~H. 1981, Journal of Geophysical Research, 86, 2261

\bibitem[{Stanislawska {et~al.}(2018)Stanislawska, Gulyaeva, Grynyshyna-Poliuga, \& Pustovalova}]{Stanislawska2018}
Stanislawska, I., Gulyaeva, T.~L., Grynyshyna-Poliuga, O., \& Pustovalova, L.~V. 2018, Space Weather, 16, 2068

\bibitem[{Strangeway {et~al.}(2005)Strangeway, Ergun, Su, Carlson, \& Elphic}]{Strangeway2005}
Strangeway, R.~J., Ergun, R.~E., Su, Y.-J., Carlson, C.~W., \& Elphic, R.~C. 2005, Journal of Geophysical Research, 110, A03221

\bibitem[{Sun {et~al.}(2025{\natexlab{a}})Sun, Zhang, Artemyev, Lu, Li, Mei, Xiang, \& O’Brien}]{Sun2025a}
Sun, W., Zhang, X.-J., Artemyev, A.~V., {et~al.} 2025{\natexlab{a}}, Journal of Geophysical Research: Space Physics, 130, e2025JA034181, \dodoi{https://doi.org/10.1029/2025JA034181}

\bibitem[{Sun {et~al.}(2025{\natexlab{b}})Sun, Zhang, Artemyev, Nakamura, \& Angelopoulos}]{Sun2025b}
Sun, W., Zhang, X.-J., Artemyev, A.~V., Nakamura, R., \& Angelopoulos, V. 2025{\natexlab{b}}, Journal of Geophysical Research: Space Physics, 130, e2025JA033882, \dodoi{https://doi.org/10.1029/2025JA033882}

\bibitem[{Sur {et~al.}(2025)Sur, Robinson, \& Garcia-Sage}]{Sur2025}
Sur, D., Robinson, R., \& Garcia-Sage, K. 2025, Space Weather, 23, e2024SW004023

\bibitem[{Thomas {et~al.}(2024)Thomas, Weigel, Pulkkinen, Schuck, Welling, \& Ngwira}]{Thomas2024}
Thomas, D., Weigel, R.~S., Pulkkinen, A., {et~al.} 2024, Journal of Geophysical Research: Space Physics, 129, e2024JA032556, \dodoi{10.1029/2024JA032556}

\bibitem[{Toffoletto {et~al.}(2003)Toffoletto, Sazykin, Spiro, \& Wolf}]{Toffoletto2003}
Toffoletto, F., Sazykin, S., Spiro, R., \& Wolf, R. 2003, Space Science Reviews, 107, 175

\bibitem[{T{\'o}th {et~al.}(2005)T{\'o}th, Sokolov, Gombosi, Chesney, Clauer, De~Zeeuw, Hansen, Kane, Manchester, Oehmke, Powell, Ridley, Roussev, Stout, Volberg, Wolf, Sazykin, Chan, Yu, \& K{\'o}ta}]{Toth2005}
T{\'o}th, G., Sokolov, I.~V., Gombosi, T.~I., {et~al.} 2005, Journal of Geophysical Research, 110, A12226

\bibitem[{T{\'o}th {et~al.}(2012)T{\'o}th, van~der Holst, Sokolov, De~Zeeuw, Gombosi, Fang, Manchester, Meng, Najib, Powell, Stout, Glocer, Ma, \& Opher}]{Toth2012}
T{\'o}th, G., van~der Holst, B., Sokolov, I.~V., {et~al.} 2012, Journal of Computational Physics, 231, 870

\bibitem[{Trondsen {et~al.}(1999)Trondsen, Lyatsky, Cogger, \& Murphree}]{Trondsen1999}
Trondsen, T.~S., Lyatsky, W., Cogger, L.~L., \& Murphree, J.~S. 1999, Journal of Atmospheric and Solar-Terrestrial Physics, 61, 829

\bibitem[{Troshichev {et~al.}(2011{\natexlab{a}})Troshichev, Sormakov, \& Janzhura}]{Troshichev2011a}
Troshichev, O., Sormakov, D., \& Janzhura, A. 2011{\natexlab{a}}, Journal of Atmospheric and Solar-Terrestrial Physics, 73, 611, \dodoi{10.1016/j.jastp.2010.12.015}

\bibitem[{Troshichev {et~al.}(2011{\natexlab{b}})Troshichev, Podorozhkina, \& Janzhura}]{Troshichev2011b}
Troshichev, O.~A., Podorozhkina, N.~A., \& Janzhura, A.~S. 2011{\natexlab{b}}, Journal of Atmospheric and Solar-Terrestrial Physics, 73, 2373, \dodoi{10.1016/j.jastp.2011.08.003}

\bibitem[{Tsurutani {et~al.}(1995)Tsurutani, Gonzalez, Gonzalez, Tang, Arballo, \& Okada}]{Tsurutani1995}
Tsurutani, B.~T., Gonzalez, W.~D., Gonzalez, A. L.~C., {et~al.} 1995, Journal of Geophysical Research, 100, 21717

\bibitem[{Tsurutani {et~al.}(2003)Tsurutani, Gonzalez, Lakhina, \& Alex}]{Tsurutani2003}
Tsurutani, B.~T., Gonzalez, W.~D., Lakhina, G.~S., \& Alex, S. 2003, Journal of Geophysical Research, 108, 1268

\bibitem[{Tsurutani {et~al.}(2006)Tsurutani, Mannucci, Iijima, Guarnieri, Gonzalez, {et~al.}}]{Tsurutani2006}
Tsurutani, B.~T., Mannucci, A.~J., Iijima, B., {et~al.} 2006, Advances in Space Research, 37, 1583

\bibitem[{Turner {et~al.}(2000)Turner, Baker, Pulkkinen, \& McPherron}]{Turner2000}
Turner, N.~E., Baker, D.~N., Pulkkinen, T.~I., \& McPherron, R.~L. 2000, Journal of Geophysical Research: Space Physics, 105, 5431, \dodoi{10.1029/1999JA000248}

\bibitem[{Turner {et~al.}(2001)Turner, Baker, Pulkkinen, Roeder, Fennell, \& Jordanova}]{Turner2001}
Turner, N.~E., Baker, D.~N., Pulkkinen, T.~I., {et~al.} 2001, Journal of Geophysical Research, 106, 19149, \dodoi{10.1029/2000JA003025}

\bibitem[{Welling(2019)}]{Welling2019}
Welling, D. 2019, in Geomagnetically Induced Currents from the Sun to the Power Grid, ed. J.~L. Gannon, A.~Swidinsky, \& Z.~Xu (American Geophysical Union (AGU)), \dodoi{10.1002/9781119434412.ch3}

\bibitem[{White {et~al.}(1998)White, Siscoe, Erickson, Kaymaz, Maynard, Siebert, Sonnerup, \& Weimer}]{White1998}
White, W.~W., Siscoe, G.~L., Erickson, G.~M., {et~al.} 1998, Geophysical Research Letters, 25, 1605, \dodoi{10.1029/98GL50865}

\bibitem[{Wilde {et~al.}(2001)Wilde, Valley, Peck, \& Graham}]{Wilde2001}
Wilde, S.~A., Valley, J.~W., Peck, W.~H., \& Graham, C.~M. 2001, Nature, 409, 175

\bibitem[{Wolf(1970)}]{Wolf1970}
Wolf, R.~A. 1970, Journal of Geophysical Research, 75, 4677

\bibitem[{Wolf {et~al.}(1982)Wolf, Harel, Spiro, Voigt, Reiff, \& Chen}]{Wolf1982}
Wolf, R.~A., Harel, M., Spiro, R.~W., {et~al.} 1982, Journal of Geophysical Research, 87, 5949

\bibitem[{Wolf {et~al.}(2007)Wolf, Spiro, Sazykin, \& Toffoletto}]{Wolf2007}
Wolf, R.~A., Spiro, R.~W., Sazykin, S., \& Toffoletto, F.~R. 2007, Journal of Atmospheric and Solar-Terrestrial Physics, 69, 288

\bibitem[{Yang {et~al.}(2008)Yang, Toffoletto, Wolf, Sazykin, Spiro, Brandt, Henderson, \& Frey}]{Yang2008}
Yang, J., Toffoletto, F.~R., Wolf, R.~A., {et~al.} 2008, Journal of Geophysical Research: Space Physics, 113, \dodoi{https://doi.org/10.1029/2008JA013635}

\bibitem[{Yu {et~al.}(2018)Yu, Yuan, Li, Huang, Wang, Yao, Funsten, \& Wygant}]{Yu2018}
Yu, X., Yuan, Z., Li, H., {et~al.} 2018, Geophysical Research Letters, 45, 8755, \dodoi{10.1029/2018GL078849}

\bibitem[{Zhang {et~al.}(2023)Zhang, Lu, \& Wang}]{Zhang2023}
Zhang, H.~X., Lu, J.~Y., \& Wang, M. 2023, Scientific Reports, 13, 7409, \dodoi{10.1038/s41598-023-34082-2}

\bibitem[{Zhang {et~al.}(2009)Zhang, Wolf, Spiro, Erickson, Sazykin, Toffoletto, \& Yang}]{Zhang2009}
Zhang, J.-C., Wolf, R.~A., Spiro, R.~W., {et~al.} 2009, Journal of Geophysical Research, 114, A08219

\end{thebibliography}

\end{document}